\documentclass[twocolumn,showpacs,preprintnumbers,amsmath,prd,amssymb,superscriptaddress,chaproman]{revtex4-1}

\usepackage{graphicx}
\usepackage{dcolumn}
\usepackage{bm}
\usepackage{multirow}
\usepackage{color}%
\usepackage{epsfig}%
\usepackage{hyperref}
\hypersetup{
colorlinks=true,
citecolor=blue,
linkcolor=blue,
urlcolor=blue
}
\usepackage[english]{babel}
\usepackage{amsmath,amssymb}
\usepackage{amsfonts}
\usepackage{stmaryrd}
\usepackage{mathrsfs}
\usepackage{xcolor}
\usepackage{color}

\newcommand{\Eq}{Eq.~}

\def\nuebar{{\rm \bar{\nu}_e}}
\def\nuebare{{\rm \bar{\nu}_{e}-e}}
\def\nue{{\rm \nu_e}}
\def\nuee{{\rm \nu_{e}-e}}
\def\s2tw{{\rm sin ^2 \theta_{W}}}

\begin{document}

\title{Constraints on nonstandard intermediate boson exchange
models from neutrino-electron scattering }

\newcommand{\deu}{Department of Physics,
Dokuz Eyl\"{u}l University, Buca, \.{I}zmir TR35160, Turkey}
\newcommand{\as}{Institute of Physics, Academia Sinica, Taipei 115, Taiwan}
\newcommand{\metu}{Department of Physics,
Middle East Technical University, Ankara TR06800, Turkey}
\newcommand{\ktu}{Department of Physics,
Karadeniz Technical University, Trabzon, TR61080, Turkey}
\newcommand{\bhu}{Department of Physics, Banaras Hindu University, Varanasi 221005, India}
\newcommand{\corr}{muhammed.deniz@deu.edu.tr}

\author{ B.~Sevda }  \affiliation{ \deu } \affiliation{ \as }
\author{ A.~\c{S}en }  \affiliation{ \deu }
\author{ M.~Demirci }  \affiliation{ \ktu }
\author{ M.~Deniz }  \altaffiliation[Corresponding Author: ]{ \corr }
\affiliation{ \deu } \affiliation{ \as }
\author{ M.~Agartioglu}  \affiliation{ \deu } \affiliation{ \as}
\author{ A.~Ajjaq }  \affiliation{ \deu }
\author{ S.~Kerman }  \affiliation{ \deu } \affiliation{ \as}
\author{ L.~Singh }  \affiliation{ \as } \affiliation{ \bhu}
\author{ A.~Sonay }  \affiliation{ \deu } \affiliation{ \as}
\author{ H.T.~Wong } \affiliation{ \as }
\author{ M.~Zeyrek }  \affiliation{ \metu }


\begin{abstract}

Constraints on couplings of several beyond-Standard-Model-physics scenarios,
mediated by massive intermediate particles
including (1) an extra Z-prime, (2) a new light spin-1 boson, and
(3) a charged Higgs boson, are placed via the neutrino-electron
scattering channel to test the Standard Model at a low energy-momentum
transfer regime. Data on $\nuebare$ and $\nuee$
scattering from the TEXONO and LSND, respectively, are used.
Upper bounds to coupling constants of the flavor-conserving and flavor-violating
new light spin-1 boson and the charged Higgs boson with respect to different
mediator masses are determined. The relevant parameter spaces are extended
by allowing light mediators. New lower mass limits
for extra Z-prime gauge boson models are also placed.

\end{abstract}


\maketitle

\section{INTRODUCTION}

Recent discovery of nonzero neutrino mass and mixing
undoubtedly implies new physics beyond the Standard
Model (BSM). Nevertheless, the origin
of the masses of neutrinos and absolute mass scale
remain unknown. The seesaw mechanisms, R-parity-violating
supersymmetry (SUSY), TeV scale loop mechanisms,
extra dimensions, and string theory are the most
popular proposals attempting to answer these questions
and explain the origin of neutrino mass~\cite{king2015}.
However, in the underlying new physics BSM, it is mostly
expected that the structure of electroweak charged and neutral
currents of the Standard Model (SM) would also change.
Such changes in the neutrino sector lead to nonstandard
interactions (NSI) of neutrinos. In many works on NSI,
new interactions are generally mediated by new particles,
which are assumed to be heavier than the electroweak scale.
Hence, these are carried out in the form of effective four-fermion
interaction at low energy. Furthermore, it is also possible
mediated new particles can have relatively low masses.

Neutrino interactions, being pure leptonic processes,
are one of the most appropriate mechanisms to test the
electroweak theory of the
SM~\cite{jpanman-wjmarciano,ohlsson2013,miranda2015,jerler125}.
Therefore, the $\nu_e(\bar{\nu}_e)-e$ elastic scattering can be used
to search for BSM new physics scenarios mediated by a massive
intermediate boson such as the extra Z-prime gauge boson ($Z'$),
new light spin-1 boson (NLS1B), and charged Higgs boson (CHB),
which are predicted by certain models to describe new
interactions mediated by new particles in addition to the
SM electroweak $Z$ and $W$ gauge bosons.

In this paper, we report experimental constraints on the masses
and coupling parameters for the exchange of the NLS1B and CHB as
well as lower mass limits on $Z'$ from  $\nuee$ and $\nuebare$
elastic scattering.

\section{NEUTRINO-ELECTRON SCATTERING AND DATA}

\subsection{Standard Model}

Since incoming neutrinos are of electron-type flavor,
$\nue(\nuebar)-e$ elastic scattering can occur via both
charged current and neutral current interaction.
Therefore, their interference, which is destructive, also contributes
to the cross section. The SM differential
cross section of $\nu_e(\bar{\nu}_e) - e$ elastic
scattering can be expressed in the laboratory frame
as~\cite{mdeniz033004,sbilmis073011,jiunn011301,sbilmis033009,bkayser87}
\begin{eqnarray}
\left[ \frac{d\sigma}{dT}(^(\bar{\nu}^)_{e}e) \right]_\text{SM}  &=&
\frac{2G_{F}^{2}m_{e}}{\pi } \left[a^{2}
+b^{2}\left(1- \frac{T}{E_{\nu }}\right) ^{2}\right. \nonumber  \\
&-& \left.ab\frac{m_{e}T} {E_{\nu}^{2}} ~ \right]~,
\label{eq::nue_gr_gl}
\end{eqnarray}
where $G_F$ is the Fermi coupling constant, $T$ is the kinetic
energy of the recoil electron, $E_\nu$ is the incident
neutrino energy, and the coefficients of $a$ and $b$
are given in Table~\ref{tab::a-b} in terms of chiral coupling
constants $g_R$ and $g_L$, weak mixing angle $sin^2{\theta_W}$, and
vector-axial-vector coupling constants $g_V$ and $g_A$, which are
defined as $-1/2+2 sin^2{\theta_W}$ and $-1/2$, respectively.

\begin{table}
\caption{Coefficients in the expression of the SM differential cross section
of $\nuebare$ and $\nuee$ scattering given in Eq.(~\ref{eq::nue_gr_gl}).}
\label{tab::a-b}
\begin{ruledtabular}
\begin{tabular}{ccc}

Coefficients    & $\bar{\nu}_{e}-e$        &  $\nu _{e}-e$              \\  \hline \hline \\
                & $(g_{V}-g_{A})/2$        &  $(g_{V}+g_{A}+2)/2$       \\ \\
    $a$         & $sin^{2}\theta _{W}$     &  $sin^{2}\theta _{W}+1/2$  \\ \\
                & $g_{R}$                  & $g_{L}+1$                  \\
                & $(g_{V}+g_{A}+2)/2$      & $(g_{V}-g_{A})/2$          \\ \\
    $b$         & $sin^{2}\theta _{W}+1/2$ & $sin^{2}\theta _{W}$       \\ \\
                & $g_{L}+1$                & $g_{R}$                    \\
\end{tabular}
\end{ruledtabular}
\end{table}

\subsection{Input data}

The analysis of the experimental data sets of TEXONO,
as a sample of the antineutrino channel at low energy
with three different detectors located at KSNL
whose energy ranges are different, and LSND, as a
sample of the neutrino channel at high energy, is reported. The published
results of the differential cross section measurements are used for each sample. The
results from three independent data sets of the TEXONO $\nuebare$
interaction are compared with those from the LSND $\nuee$ interaction.

\begin{description}

\item[TEXONO Experiment] Three experimental data sets taken with
different detectors are used as follows:
\begin{description}

\item[\bf CsI(Tl) ] 29882/7369 kg-days of reactor on/off data:
$\bar{\nu}_e-e^-$ electroweak interaction cross section,
$g_V$, $g_A$, weak mixing angle $sin^2{\theta_W}$, and
charge radius squared were measured with an effective mass
of 187 kg CsI(Tl) crystal scintillator array
at 3 $-$ 8 MeV$_{ee}$. The root-mean-square (RMS) energy resolutions
are 5.8\%, 5.2\%, and 4.0\% at $^{137}$Cs, $^{40}$K, and $^{208}$Tl
$\gamma$ peaks, respectively. The residual reactor on $-$ reactor off
event rate spectrum at 3 $-$ 8 MeV$_{ee}$ shown in Fig.~16(b) of
Ref. ~\cite{mdeniz072001} is used for this analysis.

\item[\bf HP-Ge ] 570.7/127.8 kg-days of reactor on/off data:
New limits are set to the neutrino magnetic moment and axion
with a target mass of 1.06 kg high-purity germanium
detector~\cite{hbli-htwong} at 12 $-$ 64 keV$_{ee}$.
The RMS energy resolution of HP-Ge is
880 keV$_{ee}$ at Ga-K shell x-ray energy~\cite{soma2016}.
The residual reactor on $-$ reactor off event rate spectrum at
12 $-$ 64 keV$_{ee}$ shown in Fig.~13 of
Ref. ~\cite{hbli-htwong} is used for this analysis.

\item[\bf PC-Ge ] 124.2/70.3 kg-days of reactor on/off data:
New limits are set to neutrino millicharge and low mass
weakly interacting massive particle (WIMP)
with a fiducial mass of 500 g point contact germanium (PC-Ge)
detector~\cite{jiunn011301} at the 0.3 $-$ 12 keV$_{ee}$ energy region.
The RMS energy resolution of PC-Ge is
87 keV$_{ee}$ at Ga-K shell x-ray energy~\cite{soma2016}.
The residual reactor on $-$ reactor off event rate spectrum at
0.3 $-$ 12 keV$_{ee}$ shown in Fig.~4 of
Ref. ~\cite{jiunn011301} is used for this analysis.
\end{description}

\item[\bf LSND Experiment] The Liquid Scintillator Detector at
the Los Alamos Neutron Science Center was exposed to electron
neutrinos produced at the proton beam stop with electron
recoil energy $T$ of 18 $-$ 50
MeV$_{ee}$. The cross section for the elastic scattering
reaction $\nu_e-e$ and weak mixing angle $sin^2{\theta_W}$
were measured. The energy resolution was determined
from the shape of the electron energy spectrum and
was found to be 6.6\% at the 52.8 MeV end
point. The observed and expected
distribution of beam-excess events at 18 $-$ 50
MeV$_{ee}$ published in Fig.~10 of
Ref.~\cite{auerbach2001}
are adopted in this analysis.

\item[\bf LAMPF $\nuee$ Experiment] A 15 ton fine-grained tracking calorimeter
surrounded by multiwire proportional chambers (MWPCs) was
exposed to electron neutrinos from muon decay at rest
with $T$ of $~7-60$ MeV$_{ee}$ at the Los Alamos Meson Physics Facility,
now renamed the Los Alamos Neutron Science Center.
In this experiment, neutrino-electron elastic scattering was observed and
electroweak parameters were measured. From the agreement between the measured
and SM expectation, limits on neutrino properties (such as
neutrino flavor changing neutral currents and neutrino electromagnetic
moments) and limits on the masses of new bosons [such as neutral tensor and
pseudo(scalar) boson, charged Higgs boson, and a purely left-handed
charged (neutral) vector boson] were derived in Ref.~\cite{allen93}.
\end{description}

\subsection{Analysis methods}

The expected event rate of $R$ can be calculated as
\begin{equation}
R_{X} ~ = ~ \rho_e ~ \int_{T} \int_{E_{\nu }}
\left[\frac{d\sigma}{dT}\right]'_{X}  ~ \frac{d \phi ( \nuebar ) }{dE_{\nu}} ~
dE_{\nu} ~ dT ~ ~,
\label{eq::RX}
\end{equation}
where $\rho_e$ is the electron number density per kg of target mass,
and $d\phi/dE_\nu$ is the neutrino spectrum.
$X$ represents different interaction channels such as
SM, NLS1B, CHB, etc.

The measurable differential cross section is denoted by
$\left[d\sigma/dT\right]'$ and corresponds to a convolution
of the detector energy resolution to the physical differential
cross section $\left[d\sigma/dT\right]$.
In practice, as far as BSM scenarios and experimental data
studied in this work are concerned, the variations of
$\left[d\sigma/dT\right]$ with energy are gradual, so that
the resolution smearing does not significantly alter the measured spectra
in the region of interest. The difference between $\left[d\sigma/dT\right]$
and $\left[d\sigma/dT\right]'$  is less than  0.1\%. Accordingly,
resolution effects can be neglected in this analysis.
$R_{\text{expt}}$ is expressed in units of
$\mbox{kg}^{-1} \mbox{MeV}^{-1} \mbox{day}^{-1}$
and $\mbox{kg}^{-1} \mbox{keV}^{-1} \mbox{day}^{-1}$
for the CsI(Tl) and Ge data sets, respectively.

The published neutrino
spectra for $\nu_e,~\nu_\mu,~\bar{\nu}_\mu$~\cite{auerbach2001}
are used to derive the SM differential cross sections
for the LSND analysis. The number of measured physical and
background events are taken from Fig.~10 and Table~III
of Ref.~\cite{auerbach2001}. The published total cross section
measured values are used for normalization:
\begin{eqnarray}
\sigma_{expt} &=& [10.1 \pm 1.1 (stat) \pm 1.0 (sys)] \times E_{\nu_e}
\times 10^{-45} \text{cm}^2 \nonumber \\
\sigma_{SM} &=& 9.3 \times E_{\nu_e} \times 10^{-45} \text{cm}^2.
\label{eq::lsnd}
\end{eqnarray}

The results on physics couplings from this analysis are
expressed either as ``best-fit $\pm$ statistical $\pm$
systematic uncertainties" at the 1~$\sigma$ level, or in terms of
limits at a 90\% or 95\% confidence level (C.L.).
The statistical uncertainties are derived
by the minimum $\chi^2$ method, defined as
\begin{equation}
\chi^2 = \sum\limits_{i=1} {\left[\frac{R_{expt}(i)-R_{SM}(i)-R_{X}(i)}
{\Delta(i)}\right]}^2 ~,
\label{eq::chi}
\end{equation}
where $R_{\text{expt}}$ is the measured rate; and $R_{\text{SM}}$ and $R_{X}$ are the expected
event rates for the SM and $X$ (with $X = Z'$, NLS1B, CHB, etc.), respectively;
and $\Delta(i)$ is the $i$th bin statistical uncertainty
published by the experiments. The published systematic uncertainties
of the experiments contribute to shifts of the best-fit values in the
physics couplings. The two contributions are added in quadrature to give
rise to the combined uncertainties, from which the 90\% or 95\% C.L.
limits can be  derived using the prescription of Ref. ~\cite{Gary98}.

\begin{figure}
\includegraphics[width=8.0cm]{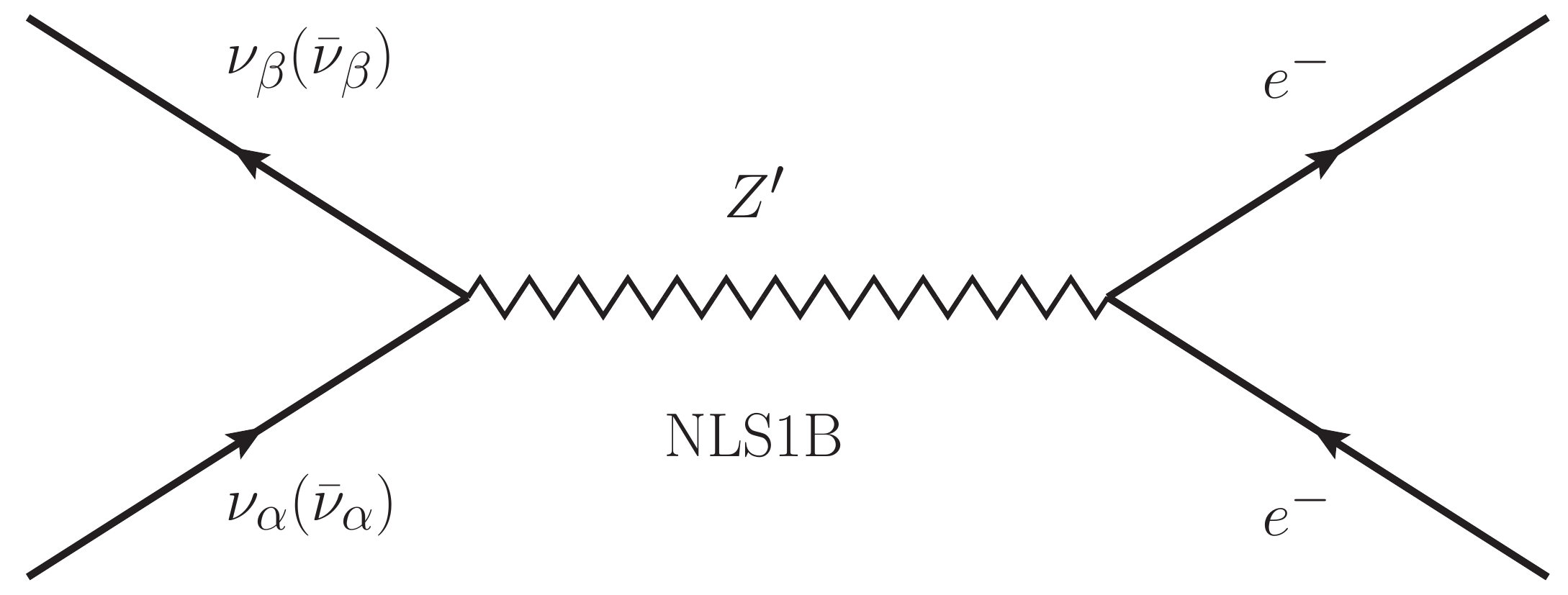}
\caption{Feynman diagram for $\nue (\nuebar)-e$ via exchange of massive mediators
such as the virtual $Z'$ or NLS1B.}
\label{fig::feydiag}
\end{figure}

\section{INTERMEDIATE BOSONS BEYOND THE STANDARD MODEL}

Some of the BSM involve exchanging of massive intermediate
bosons such as the extra $Z'$, NLS1B, and CHB
in addition to the SM $Z$ and $W$ gauge bosons.
A Feynman diagram of neutrino and antineutrino scattering
off electron for various NSI
scenarios is illustrated in Fig.~\ref{fig::feydiag}.

Some of the new physics BSMs have a mechanism giving mass to
neutrinos such as low-energy SUSY with R-parity breaking,
an extra Higgs boson, unified SUSY models, etc.
Indeed, any BSM physics model should reproduce current data and
therefore should include massive neutrinos. In addition, there are
some recent model-dependent BSM studies in the literature~\cite{ferzan,laha}.
In this paper, we only study some specific
models for new interactions with massive virtual bosons. In the following
sections these BSM scenarios and their corresponding experimental constraints
will be discussed in detail.

NSI can simply be considered as modifications of coupling
constants with additional new terms in the chiral couplings
of $g_{R,L}$ in general. Therefore, for the
flavor-conserving (FC) NSI cases, the new couplings can be expressed as
\begin{equation}
g_{R(L)} \rightarrow \tilde{g}_{R(L)}=g_{R(L)}+\tilde{\varepsilon}^{R(L)}_{ee}.
\end{equation}

\begin{table}
\caption{Coefficients with BSM contributions in expressions of the differential
cross section of $\nue(\nuebar)-e$ scattering given in Eq.~(\ref{eq::nue_gr_gl}).}
\label{tab::BSM_a-b}
\begin{ruledtabular}
\begin{tabular}{ccc}

Coefficients    & $\bar{\nu}_{e}-e$   &  $\nu _{e}-e$  \\  \hline \hline \\
    $a^2$         & $\tilde{g}_R^{2} + \sum\limits_{\ell^\prime \neq e}|\tilde{\varepsilon}^R_{e\ell^\prime}|^2$
                & $\left(\tilde{g}_L + 1 \right)^{2} + \sum\limits_{\ell^\prime \neq e}(\tilde{\varepsilon}^L_{e\ell^\prime})^2$   \\

    $b^2$       & $\left(\tilde{g}_L + 1 \right)^{2} + \sum\limits_{\ell^\prime \neq e}
                |\tilde{\varepsilon}^L_{e\ell^\prime}|^2$
                & $\tilde{g}_R^{2} + \sum\limits_{\ell^\prime \neq e}
                (\tilde{\varepsilon}^R_{e\ell^\prime})^2$  \\
    $ab$        & \multicolumn{2}{c}{$\tilde{g}_R \left(\tilde{g}_L + 1 \right)
                + \sum\limits_{\ell^\prime \neq e}|\tilde{\varepsilon}^R_{e\ell^\prime}||\tilde{\varepsilon}^L_{e\ell^\prime}|$ } \\

\end{tabular}
\end{ruledtabular}
\end{table}

The $\nuebare$ and $\nuee$
scattering differential cross sections can be written in
terms of new couplings of FC and
flavor-violating (FV) NSI of neutrinos given
in Table~\ref{tab::BSM_a-b}.
The differential cross section of BSM contributions
can be obtained by using Eq.~(\ref{eq::nue_gr_gl}) together with
the coefficients from Table~\ref{tab::BSM_a-b} considering both
FC NSI and FV NSI with $\ell^\prime=\mu$ or $\tau$.

\begin{figure}[!ht]
  \begin{center}
  \includegraphics[width=8.5cm]{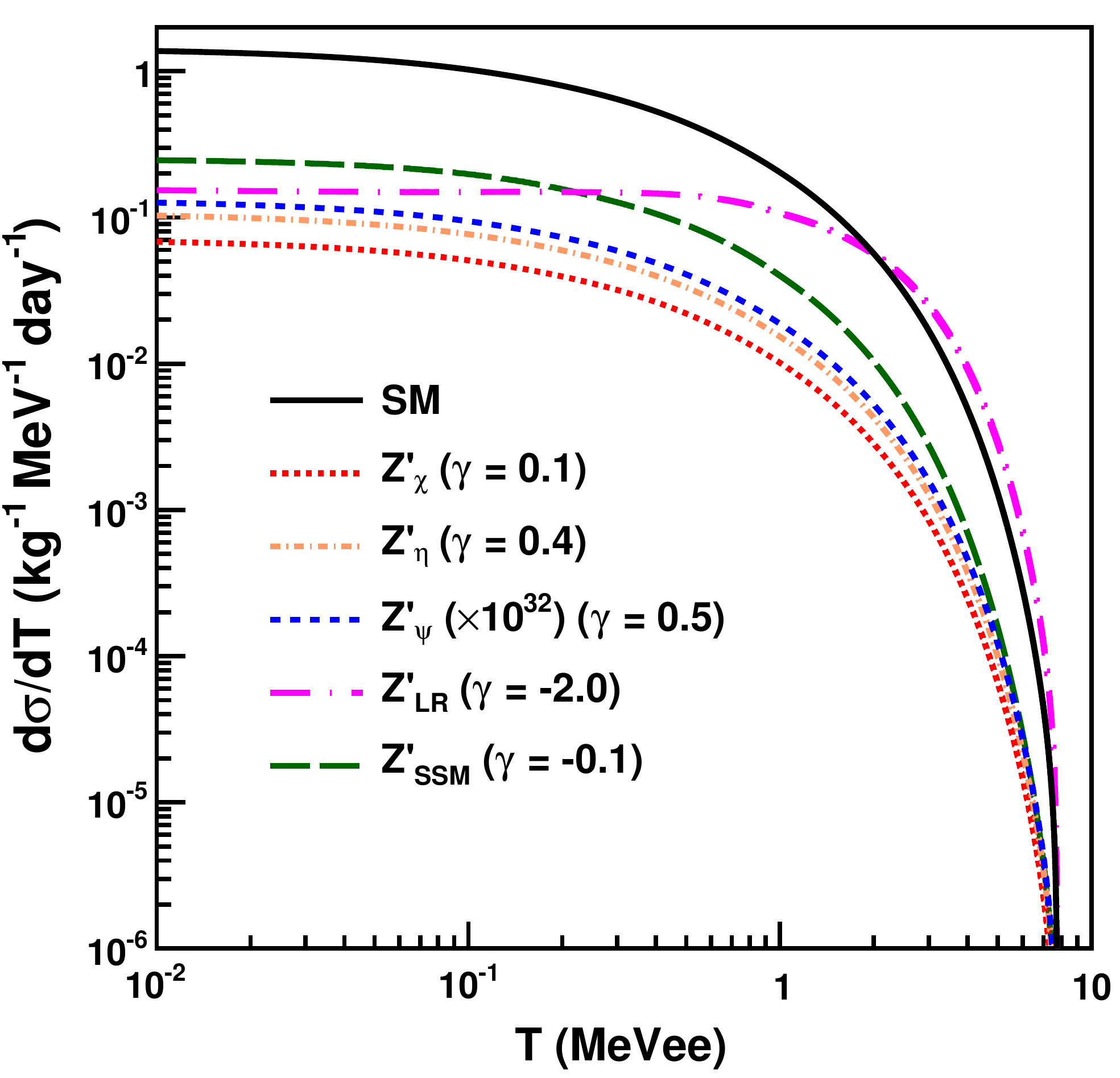}\qquad\qquad
  \caption{\label{fig::diff_z-prime} Differential cross
  section as a function of the recoil energy $T$ with typical reactor
  $\nuebar$ spectra for extra $Z'$ models for a specific $\gamma$ value
  using CsI(Tl) as a target, where $ \gamma={(M_{Z}/M_{Z'})}^2 $.
  The SM contributions are superimposed for comparison.}
  \end{center}
\end{figure}

\begin{table*} [hbt]
\caption{
Constraints on $M_{Z'}$ at 95\% C.L. obtained from the best fit on $\gamma$,
current limits, and projected sensitivities on $M_{Z'}$ bounds by improving
1\% in the accuracies of CsI(Tl) data.}
\label{tab:CsI}
\begin{ruledtabular}
\begin{tabular}{cccccccc}
\multicolumn{2}{c} {\multirow{3}{*}{Model}} & \multirow{2}{*}{Best fit} & \multirow{3}{*}{$\chi ^2_{min}$/dof}
& $ M_{Z'} $bounds & Projected (1\%) & Current limit\\
\multicolumn{2}{c}{} & \multirow{2}{*}{for $\gamma$ (1~$\sigma$)} &  & at 95\% C.L.  & $M_{Z'}$ bounds & [PDG 2016] \\
\multicolumn{2}{c}{} &     &    &  (GeV)   & at 95\% C.L. (GeV)   & at 95\% C.L. (GeV) \\ \hline \\
\multirow{3}{*}{$E_6$ String}  & ~ $Z'_{\chi}$ & $0.16 \pm 0.41 \pm 0.31$ &  8.7/9  & $ > 85 $  & $ > 915 $  &  $ > 1970 $ (ATLAS) \\ \\
\multirow{3}{*}{Type}  &  ~ $Z'_{\eta}$ & $0.43 \pm 1.01 \pm 0.83$ &  8.7/9  & $ > 52 $  & $ > 566 $ & $ > 1870 $ (ATLAS) \\ \\
& ~ $Z'_{\psi}$ & $[0.44 \pm 1.13 \pm 0.95]\times10^{-18}$ &  8.7/9  & $ > 0 $  & $ > 0 $ & $ > 2260 $ (CMS) \\ \\
\multicolumn{2}{c}{~ $Z'_{LR}$} & $-8.02 \pm 5.28 \pm 0.61$ &  7.8/9  & $ > 44 $  & $ > 413 $ & $ > 1162 $ (RVUE) \\ \\
\multicolumn{2}{c}{~ $Z'_{SSM}$} & $-0.04 \pm 0.14 \pm 0.06$ &  8.7/9  & $ > 172 $  & $ > 1822 $ & $ > 1830 $ (ATLAS) \\
\end{tabular}
\end{ruledtabular}
\end{table*}

\subsection{Extra $Z'$ gauge boson}

Intermediate particles of electroweak interaction in addition
to SM $W^{\pm}$ and $Z^0$ gauge bosons, have engaged
particle physicists' attention for a long while since they are a common
feature of many models aiming to define the nature of BSM.
The $Z'$ gauge boson, the new gauge boson, was proposed as a
theoretical particle resulting from the expansion of electroweak
interactions in particle physics. Its name comes from the SM
$Z$ boson.

New massive U(1) gauge bosons emerge in grand unified and
superstring theories such as SO(10) and $E_6$~\cite{London86}, in theories
of extra space-time dimensions of the SM gauge bosons~\cite{petriello}.
In this study, we will not restrict ourselves to SM gauge bosons.
In fact, we will consider a possible new vector boson predicted
in many extensions of the SM called the $Z'$ gauge boson,
which is a massive, electrically neutral and color-singlet
hypothetical particle of spin 1.

There are various physical models of BSM that suggest different
$Z'$ bosons. The most popular of them are the $E_6$-string-type model,
left-right symmetric model, and the sequential Standard Model (SSM).
The $E_6$-string-type model, based on $E_6$ symmetries,
contains the $SO(10) \times U(1)_{\psi}$ and $SU (5) \times U(1)_{\chi}$,
which means that the two $Z'$ states (i.e., $Z'_{\chi}$ and $Z'_{\psi}$)
are included and can mix by some angle $\beta$. The mixing of these two states
is given by their linear combination as
$Z'(\beta) = Z'_{\chi}(cos\beta)+Z'_{\psi}(sin\beta)$~\cite{erler}.

The new coupling parameters of BSM are generally
obtained by modifying the ordinary coupling constants of the SM.
Therefore, the new cross sections for the interactions via
the exchange of an extra $Z'$ gauge boson can be obtained by replacing the
SM couplings appearing in \Eq(\ref{eq::nue_gr_gl}) with the new modified
couplings accordingly.

The new differential cross section of $Z'$ models
for $\nu_e(\bar{\nu}_e)-e$ elastic scattering can be
obtained by modifying the couplings with
\begin{eqnarray}
\tilde{\varepsilon}^R _{ee}&=& 2\gamma\sin^2{\theta_W}
\rho^{NC}_{\nu e}\left(\frac{c_\beta}{2\sqrt{6}}-\frac{s_\beta}{3}
\sqrt{\frac{5}{8}}\right)\left(\frac{3c_\beta}{2\sqrt{6}}+
\frac{s_\beta}{3}\sqrt{\frac{5}{8}}\right) \nonumber \\
\tilde{\varepsilon}^L_{ee} &=& 2\gamma\sin^2{\theta_W} \rho^{NC}_{\nu e}
{\left(\frac{3c_\beta}{2\sqrt{6}}+
\frac{s_\beta}{3}\sqrt{\frac{5}{8}}\right)}^2 ~,
\label{eq::epslnl}
\end{eqnarray}
where $c_\beta=cos\beta$, $s_\beta=sin\beta$,
and $\gamma = {(M_{Z}/M_{Z'})}^2$.

In this paper, three main models of the $E_6$-string-type
model~\cite{barrancoZprime}
have been investigated: the $\chi$ model where
${cos\beta}=1$, the $\psi$ model where ${cos\beta}=0$,
and the $\eta$ model where ${cos\beta}=\sqrt{3/8}$.

One of the other popular models proposing a heavy neutral
vector boson is the left-right symmetric model, which
has breaking dynamical symmetry. The
left-right symmetric model is based on
$SU(2)_L \times SU(2)_R \times U(1)_{B-L}$,
where $SU(2)_L$ and $SU(2)_R$ are associated to the left-handed
and right-handed weak isospins, respectively, and $U(1)_{BL}$
is associated to the charge $Q_{BL} = B-L$, where $B$ and $L$ are
the baryon and lepton number, respectively.
The couplings are constructed in this model as
\begin{eqnarray}
\tilde{g}_R &=& Ag_R + Bg_L
\text{\ \ \ \ and \ \ \ \ } \nonumber \\
\tilde{g}_L &=& Ag_L + Bg_R ~,
\label{eq::grlr}
\end{eqnarray}
where the parameters of $A$ and $B$ can be described as
\begin{eqnarray}
A &=& 1+\frac{\sin^4{\theta_W}}{1-2\sin^2{\theta_W}}\gamma
\text{\ \ \ \ and \ \ \ \ } \nonumber \\
B &=& \frac{\sin^2{\theta_W}(1-\sin^2{\theta_W})}{1-2\sin^2{\theta_W}} ~.
\label{eq::ab}
\end{eqnarray}

Finally, the SSM, $Z'_{SSM}$, is defined as having the same couplings
with quarks and leptons which are identical to those of the SM $Z$,
and decays of only known fermions. This model serves as
a useful reference case when comparing the
$Z'$ researches with well-motivated models \cite{erler}.
The differential cross section for this model can be
written as
\begin{eqnarray}
\left[ \frac{d\sigma}{dT}(\bar{\nu}_{e}e )\right] _{Z'_{SSM}} &=&
\frac{2G_{F}^{2}m_{e}}{\pi }\left\{\gamma \left[4g_{L}
\left(1-\frac{T}{E_{\nu}}\right)^{2}\right. \right. \nonumber  \\
&-&\left.2g_{R}\frac{m_{e}T} {E_{\nu}^{2}}\right]
+\gamma{^2}\left[g_{R}^{2} + g_{L}^{2}
\left(1-\frac{T}{E_{\nu }}\right)^{2}\right.\nonumber  \\
&-&\left. \left.g_{R}g_{L}\frac{m_{e}T} {E_{\nu}^{2}}\right]\right\}~.
\label{eq::Z_SSM}
\end{eqnarray}

\begin{figure}[!ht]
  \begin{center}
  \includegraphics[width=8.5cm]{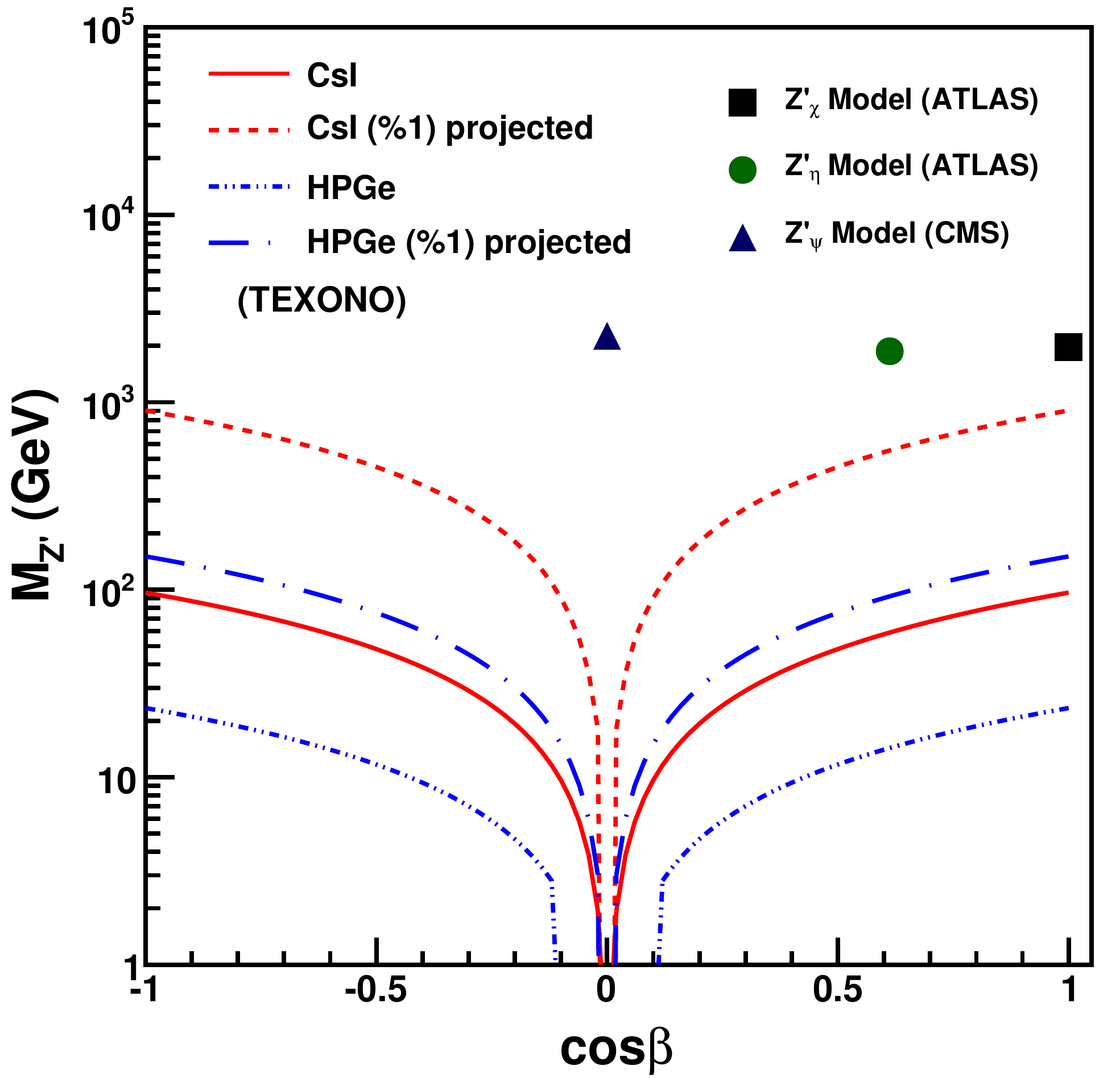}\\[4ex]
  \caption{\label{fig::cosT_csi}
  The lower limit for the mass of $Z'$ at 95\% C.L. using the TEXONO CsI(Tl)
  and HP-Ge detector data sets for the mixing-parameter-independent case of the
  E6-string-type model. Projected sensitivities by improving the
  experimental accuracies to \% 1 are superimposed.}
  \end{center}
\end{figure}

The differential cross sections for various extra $Z'$ models
with the use of CsI(Tl) as a target at a specific value of $\gamma$
are displayed in Fig.~\ref{fig::diff_z-prime}, where the SM contribution
is superimposed for comparison. As it can be seen in the figure, the
cross sections of different $Z'$ models demonstrate similar behavior
with respect to the recoil energy of the electron. Working at the MeV-energy
regime has many more advantages than working at low energy since the
cross sections of the SM were measured more precisely with CsI(Tl) data.
Therefore, more stringent limits are set to the mass of the extra $Z'$
gauge boson with the CsI(Tl) detector data set compared to those of Ge
detector data sets.

By adopting a minimum $\chi^2$ analysis, the best-fit results and
the lower bounds for the mass of the $Z'$ gauge boson obtained from
the CsI(Tl) detector data set for each $Z'$ model are given in
Table~\ref{tab:CsI}. The projected sensitivities and the present
bounds from the LHC experiment are also given for comparison.
It can be seen that the bounds from low-energy neutrino-electron
scattering experiments are much less stringent than those of high-energy
collider experiments, due to worse statistics and in general
a larger background.

\begin{figure}[!ht]
  \begin{center}
  \includegraphics[width=8.5cm]{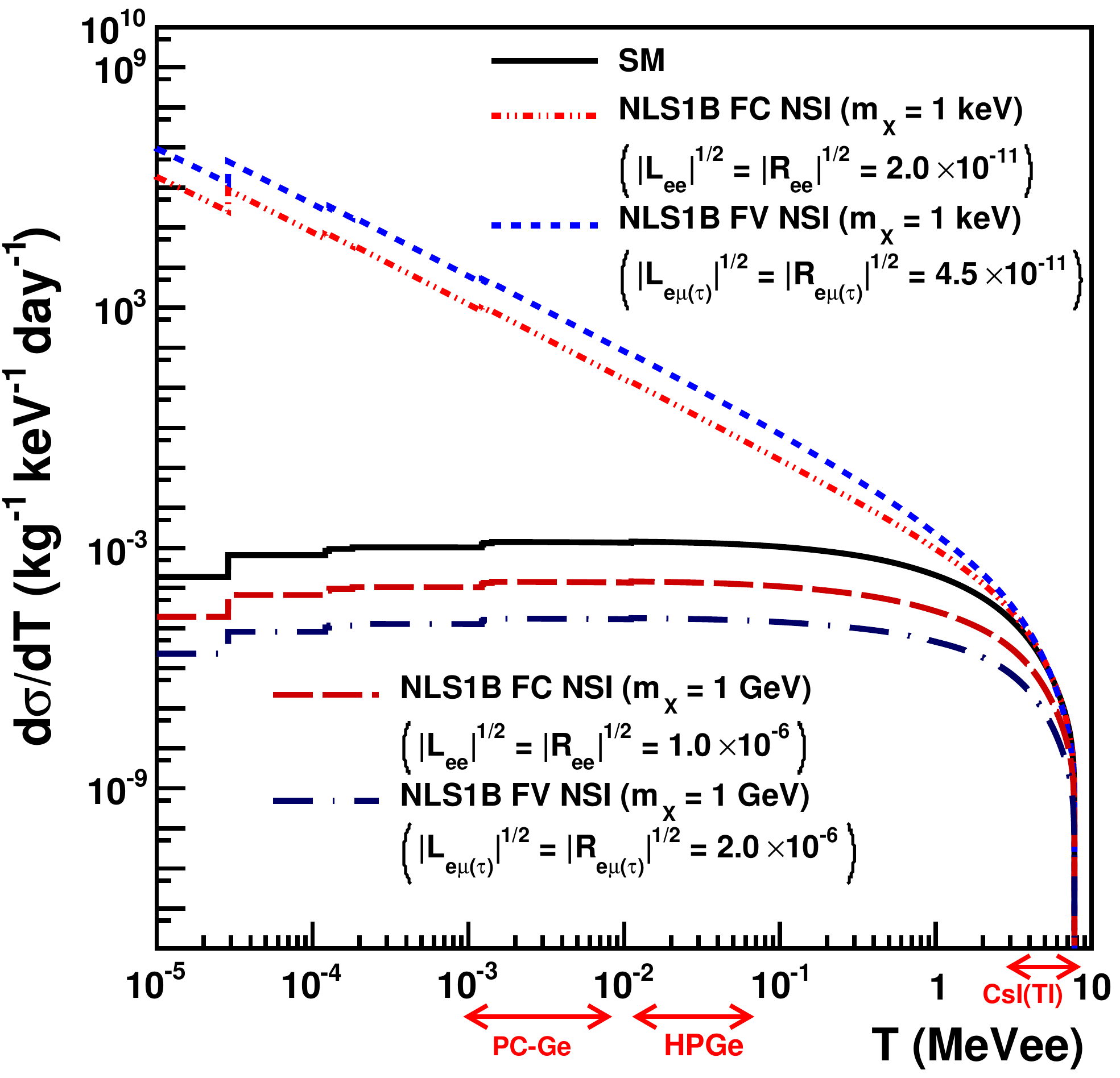}
  \caption{Differential cross section
  as a function of the recoil energy $T$ with
  typical reactor $\nuebar$ spectra
  for the NLS1B at some specific coupling and
  mass parameters relevant to this work using Ge as a target
  for both the FC and FV cases.
  The SM contributions are superimposed for comparison.}
  \label{fig::diff_nls1b}
  \end{center}
\end{figure}

The realistic sensitivities of future reactor $\nuebare$ scattering
experiments are discussed in Table~VII of Ref.~\cite{mdeniz072001}.
The main improvement is due to background suppression. The
effects of  a projected accuracy of ~1\% to the various $M_{Z'}$ bounds
are also shown in Table~\ref{tab:CsI}.
The aim of this extrapolation is to see how the $Z'$ mass bounds are related to
the experimental accuracies. It may provide intuitive scaling for
the future neutrino experiments.
Moreover, the mixing-parameter-independent sensitivities
of CsI(Tl) and HP-Ge detector data at 95\% C.L. for $E_6$-string-type
$Z'$ models are shown in Fig.~\ref{fig::cosT_csi}.
It can be seen from the figure that the $\chi$ model, where $cos\beta=1$,
can provide a more stringent limit.

\begin{table*} [hbt]
\caption{
Constraints at 90\% C.L. on the couplings for the FC NLS1B with $m_X^2\ll2m_eT$ and
$m_X^2\gg2m_eT$ for the TEXONO and LSND data sets
obtained from a one-parameter-at-a-time analysis.}
\label{tab::nls1b_fc}
\begin{ruledtabular}
\begin{tabular}{cccc}
\multicolumn{2}{c}{($m_X^2\ll2m_eT$)} & \multicolumn{2}{c}{($m_X^2\gg2m_eT$)} \\ \ \\
TEXONO PC-Ge $(\times 10^{-6})$ & LSND $(\times 10^{-6})$ &
TEXONO CsI(Tl) $(\times 10^{-6})$ & LSND $(\times 10^{-6})$ \\ \hline \\

${|L_{ee}|}^{1/2} < 1.21 $ & ${|L_{ee}|}^{1/2} < 10.12$ & ${|L_{ee}|}^{1/2}/m_X < 2.58$ & ${|L_{ee}|}^{1/2}/m_X < 1.87$ \\ \ \\

${|R_{ee}|}^{1/2} < 1.22$ & ${|R_{ee}|}^{1/2} < 31.42$ & ${|R_{ee}|}^{1/2}/m_X < 1.87$ & ${|R_{ee}|}^{1/2}/m_X < 8.63$ \\ \ \\

${|V_{ee}|}^{1/2} < 1.02$ & ${|V_{ee}|}^{1/2} < 9.95$ & ${|V_{ee}|}^{1/2}/m_X < 1.58$ & ${|V_{ee}|}^{1/2}/m_X < 1.84$ \\  \ \\

${|A_{ee}|}^{1/2} < 1.02$ & ${|A_{ee}|}^{1/2} < 10.22$ & ${|A_{ee}|}^{1/2}/m_X < 1.33$ & ${|A_{ee}|}^{1/2}/m_X < 1.88$ \\ \ \\

\end{tabular}
\end{ruledtabular}
\end{table*}

\begin{figure*}[]
  \begin{center}
  {\bf \hspace{0.5cm}(a)} {\bf \hspace{7.5cm}(b)}\\
    \includegraphics[width=7.cm]{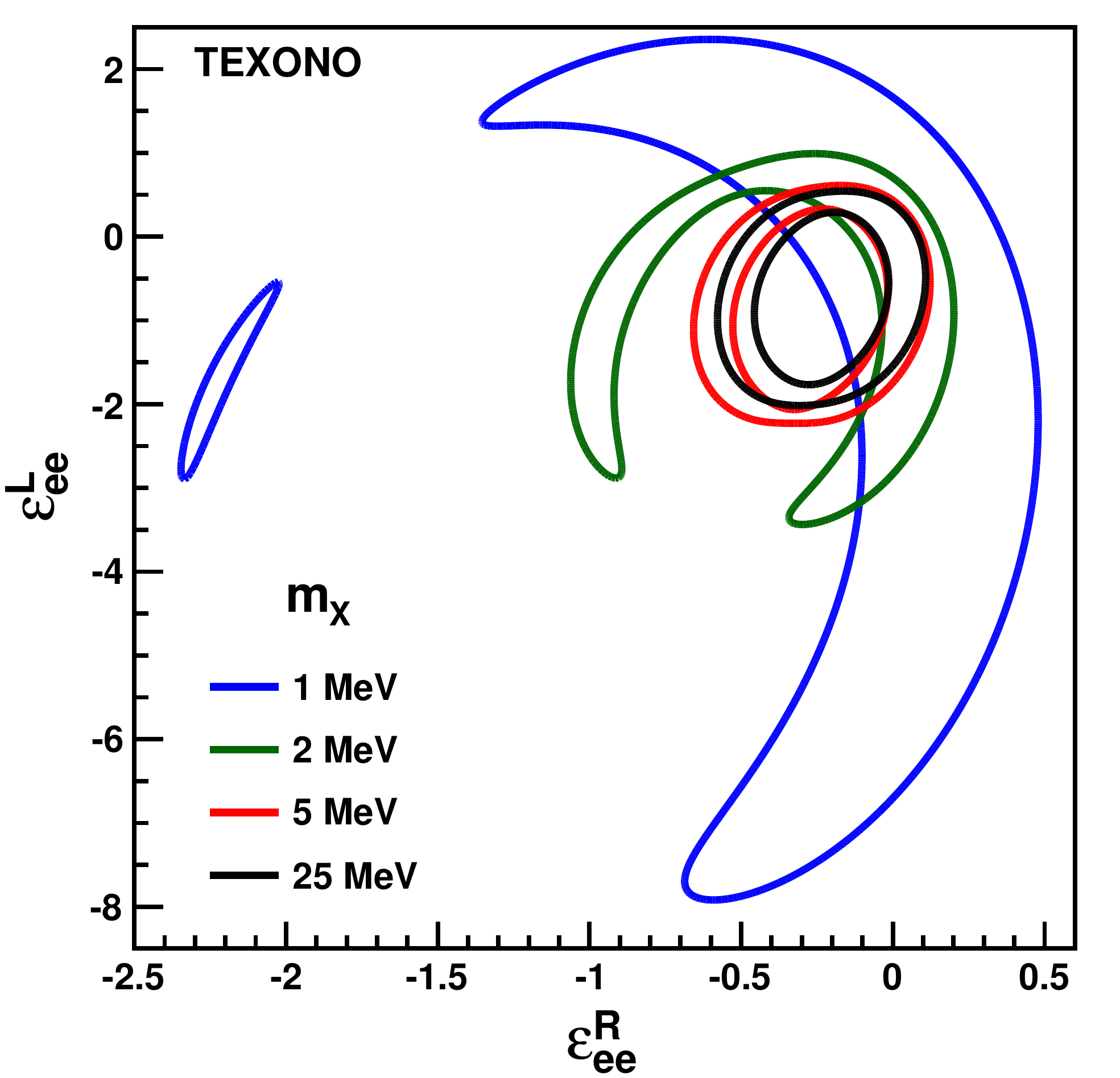} \qquad
    \includegraphics[width=7.cm]{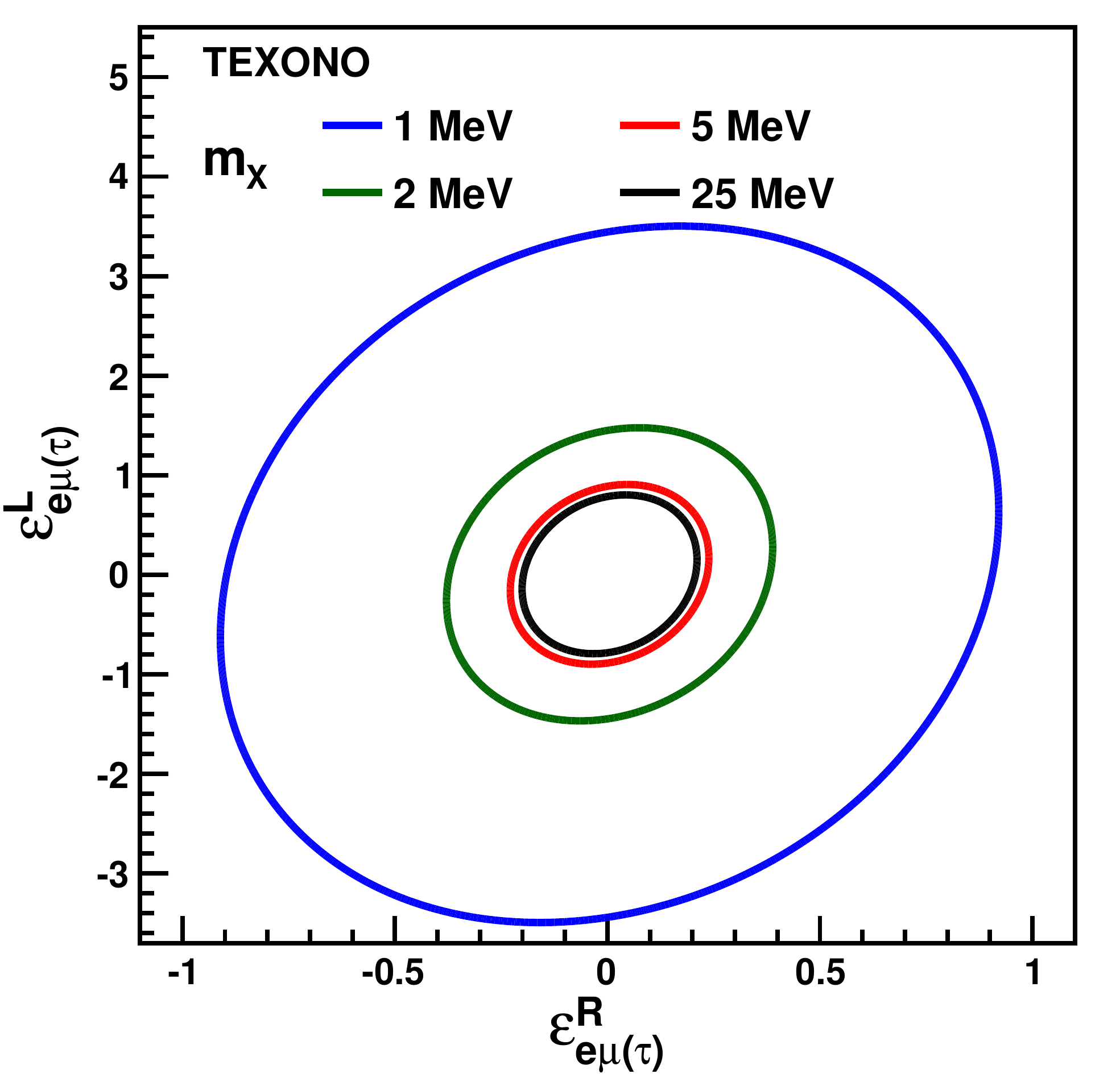} \\
{\bf \hspace{0.5cm}(c)} {\bf \hspace{7.5cm}(d)}\\
    \includegraphics[width=7.cm]{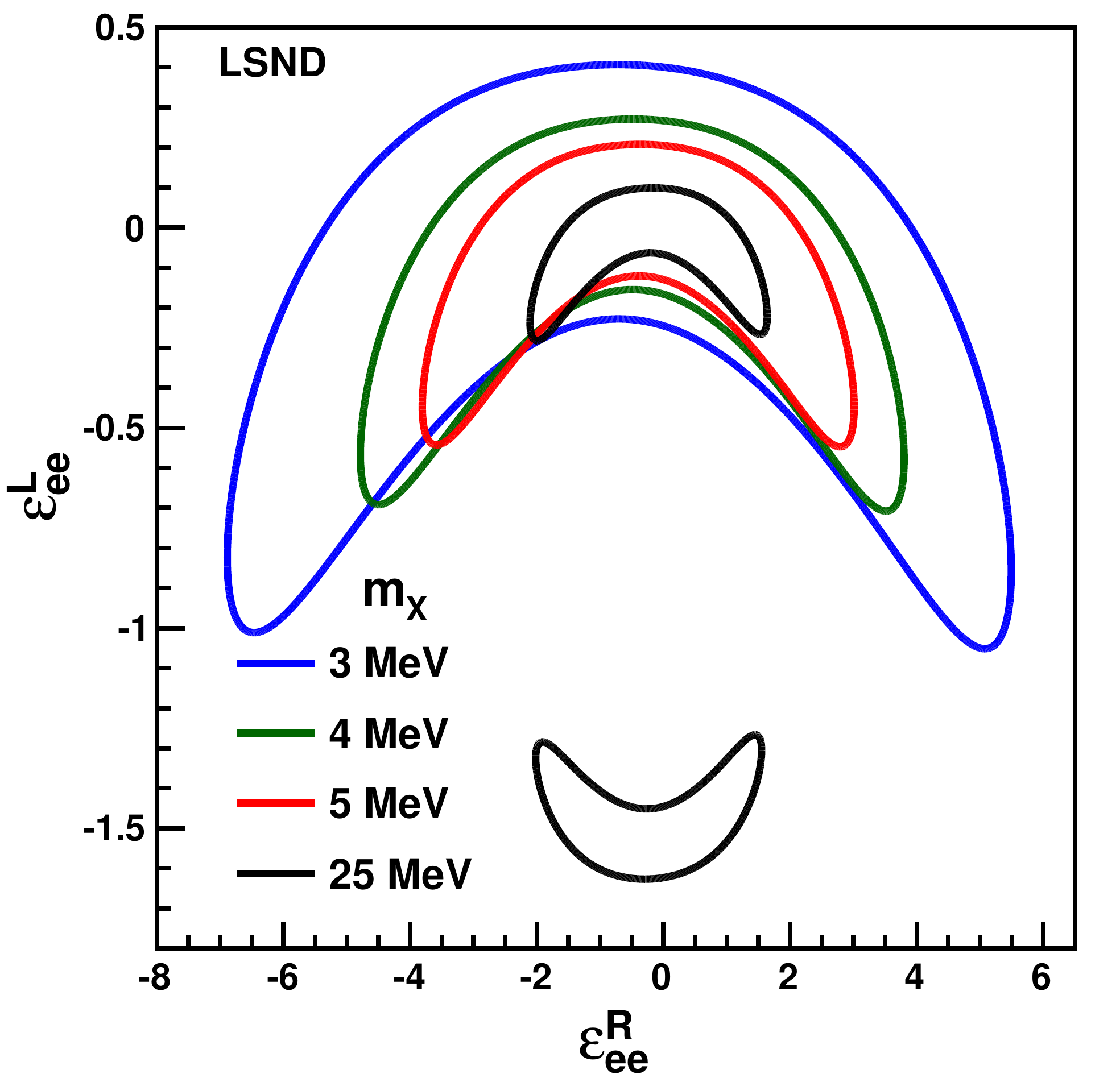} \qquad
    \includegraphics[width=7.cm]{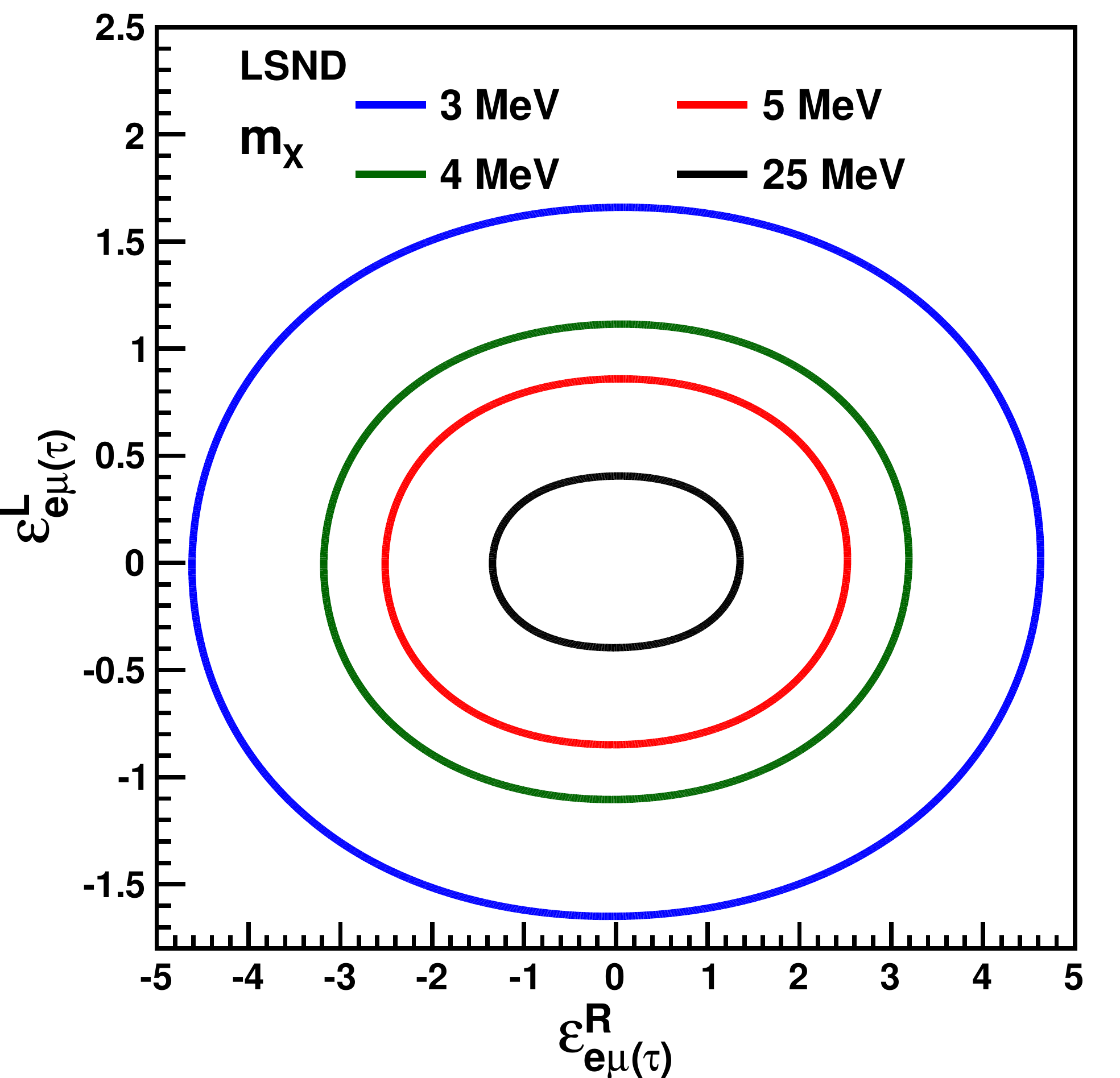}\\
    {\bf \hspace{0.5cm}(e)} {\bf \hspace{7.5cm}(f)}\\
    \includegraphics[width=7.cm]{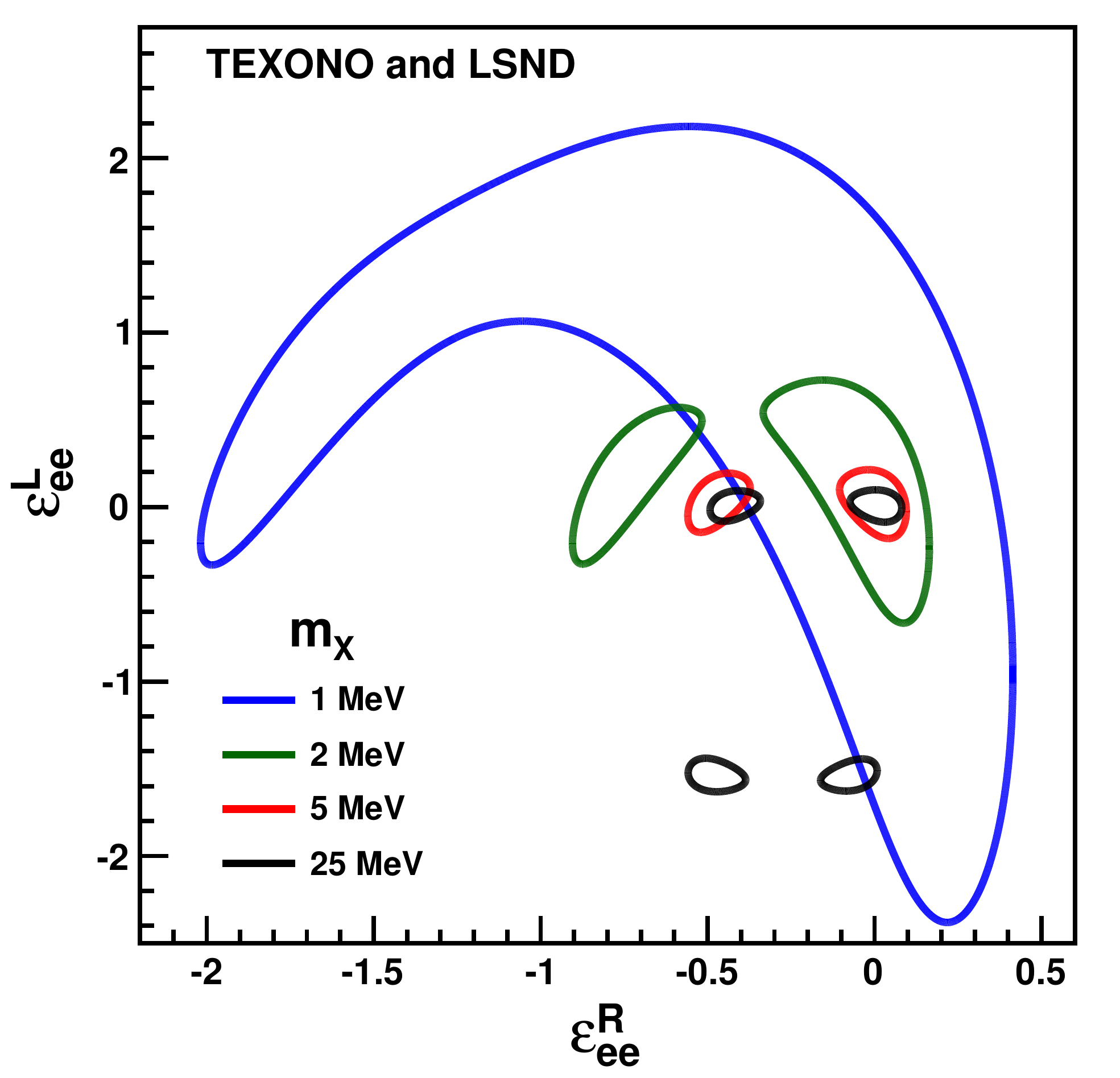}\qquad
    \includegraphics[width=7.cm]{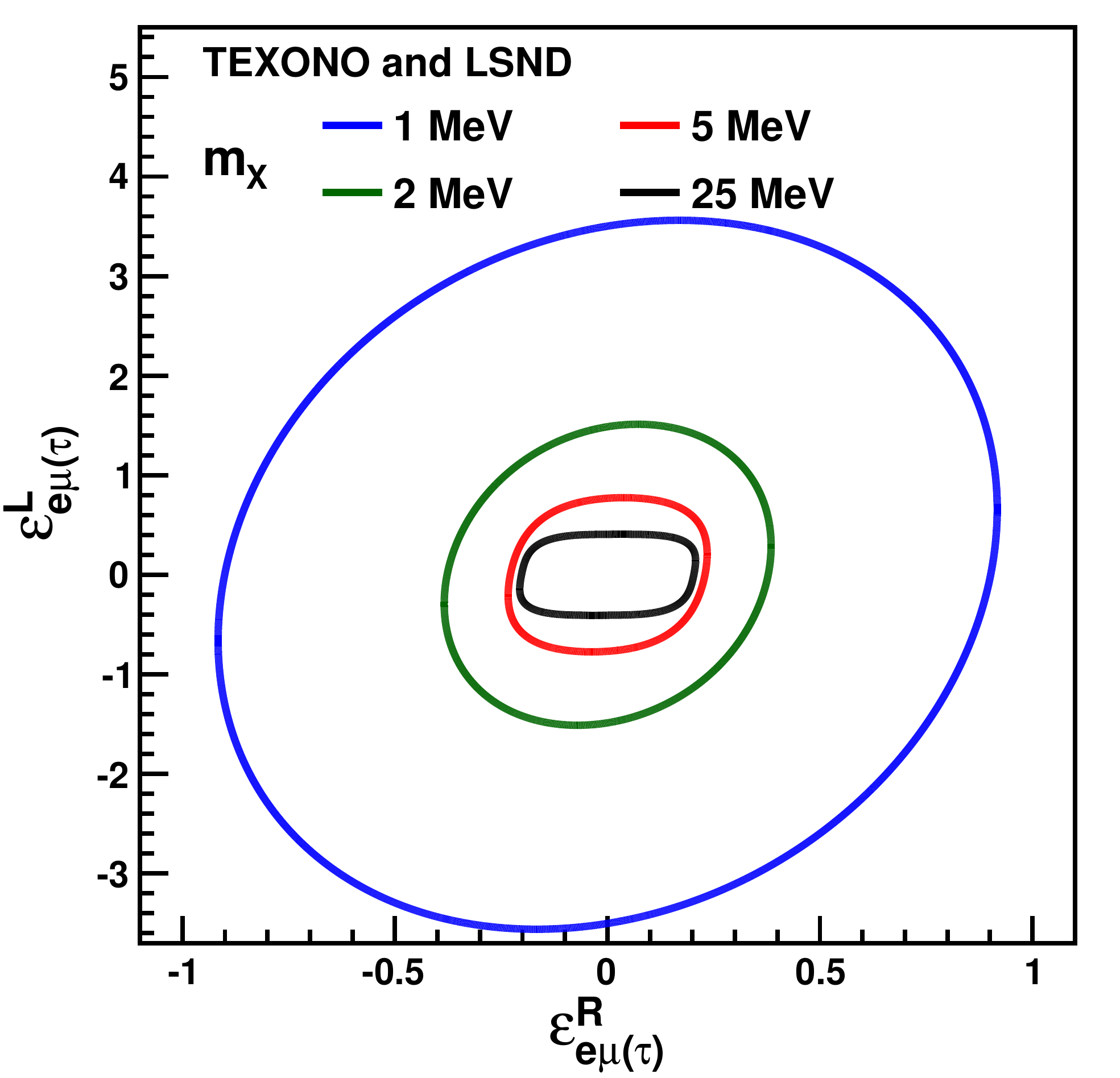}\\
    \caption{The allowed regions at 90\% C.L. for (a) the FC NLS1B in the parameter space of $\varepsilon^L_{ee}$
    and $\varepsilon^R_{ee}$; (b) FV NLS1B in the parameter space of $\varepsilon^L_{e\mu(\tau)}$
    and $\varepsilon^R_{e\mu(\tau)}$ for TEXONO CsI(Tl) with various $m_{X}=$ 1, 2, 5, 25 MeV, from
    outer to inner, respectively; (c) FC NLS1B in the parameter space of $\varepsilon^L_{ee}$
    and $\varepsilon^R_{ee}$; (d) FV NLS1B in the parameter space of $\varepsilon^L_{e\mu(\tau)}$
    and $\varepsilon^R_{e\mu(\tau)}$ for LSND with various $m_{X}=$ 3, 4, 5, 25 MeV, from
    outer to inner, respectively, with the global fitting for allowed regions of TEXONO CsI(Tl)
    and LSND at 90\% C.L. for (e) FC NLS1B couplings of $\varepsilon^L_{ee}$ vs $\varepsilon^R_{ee}$
    and (f) FV NLS1B couplings of $\varepsilon^L_{e\mu(\tau)}$
    vs $\varepsilon^R_{e\mu(\tau)}$ with various $m_{X}=$ 1, 2, 5, 25 MeV from
    outer to inner, respectively.}
       \label{fig::s1_res}
  \end{center}
\end{figure*}

\subsection{New light spin-1 boson}

The exchange of new massive particles can be a possible origin
of NSI of neutrinos, manifested as anomalies in the measurable
total or differential cross sections. These massive particles,
however, can be as light as in the order of a few MeV scale,
which is the range of low-energy experiments.
The NLS1B is one of the examples of such kinds of particles.
A spin-1 particle could also be involved in explaining the NuTeV
anomaly~\cite{Celine2004}. In addition to this, the NLS1B may also
explain the muon anomalous magnetic moment value~\cite{Gninenko2001}.
Moreover, spin-1 bosons can couple to dark matter and
the nonbaryonic matter of the Universe in the MeV scale region.
They could be responsible for the annihilation that is seen as
the unexplained 511~keV gamma emissions anomaly from the galactic
bulge~\cite{Hooper2007}. Furthermore, the NLS1B particle, which is lighter
than $b$ quarks, would explain the anomalous $CP$-violation in the mixing of
neutral B-mesons.~\cite{Sechul2011}. The effective Lagrangian for the NLS1B
can be written as~\cite{ChengWC2013}
\begin{eqnarray}
{\cal L}_X = &-& g_{\nu_\ell\nu_{\ell^\prime}}\bar\nu_\ell\gamma^\mu P_L\nu_{\ell^\prime}X_\mu \nonumber \\
&-&\bar e\gamma^\mu\left(L_{e\ell^\prime}P_L+R_{e\ell^\prime}P_R\right)eX_\mu ~,
 \label{eq::LX}
\end{eqnarray}
where $P_{R,L}=(1\pm\gamma^{5})/2$ are the chiral projectors and
the labels $\ell,\ell^\prime$ correspond to lepton flavor
$e, \mu,$ or $\tau $.

The $\nue(\nuebar) - e$ scattering differential cross section for NLS1B exchange
contributions can be obtained by modifying the chiral couplings
as given in Table~\ref{tab::BSM_a-b}. The $\tilde{\varepsilon}^{R(L)}_{e\ell^\prime}$ can be defined
in terms of the coupling parameters $R_{e\ell^\prime},~L_{e\ell^\prime}$, and the mass of
$m_X$ as
\begin{eqnarray}
\tilde{\varepsilon}^R_{e\ell^\prime}&=&\frac{R_{e\ell^\prime}}
{2\sqrt2 G_{F}(2m_e T + m_X^2)} = \frac{m_X^2}
{2m_e T + m_X^2} \varepsilon^R_{e\ell^\prime} \nonumber \\
\tilde{\varepsilon}^L_{e\ell^\prime}&=&\frac{L_{e\ell^\prime}}
{2\sqrt2 G_{F}(2m_e T + m_X^2)} = \frac{m_X^2}
{2m_e T + m_X^2} \varepsilon^L_{e\ell^\prime}~,
\label{eq::coup_tilde_nls1b}
\end{eqnarray}
where $\ell^\prime=e,\mu,$ or $\tau$, and
$\varepsilon^{R(L)}_{e\ell^\prime}$ can be defined as
\begin{eqnarray}
\varepsilon^R_{e\ell^\prime}&=&\frac{R_{e\ell^\prime}}
{2\sqrt2 G_{F}m_X^2} \nonumber \\
\varepsilon^L_{e\ell^\prime}&=&\frac{L_{e\ell^\prime}}
{2\sqrt2 G_{F}m_X^2}~.
\label{eq::coup_nls1b}
\end{eqnarray}

We can alternatively define new couplings $V(A)_{e\ell^\prime}$ and
$\tilde{\varepsilon}^{V(A)}_{e\ell^\prime}$ similar to the SM chiral couplings of
$g_{L},~g_{R}$ in the case of one of the couplings not being zero as
\begin{eqnarray}
V_{e\ell^\prime}&=&(L_{e\ell^\prime}+R_{e\ell^\prime})/2 \nonumber \\
A_{e\ell^\prime}&=&(L_{e\ell^\prime}-R_{e\ell^\prime})/2 \nonumber \\
\tilde{\varepsilon}^{V(A)}_{e\ell^\prime}&=&\frac{\tilde{\varepsilon}^L_{e\ell^\prime}
\pm \tilde{\varepsilon}^R_{e\ell^\prime}}{2}
=\frac{m_X^2}{2m_e T + m_X^2} \frac{\varepsilon^L_{e\ell^\prime} \pm \varepsilon^R_{e\ell^\prime}}{2} \nonumber \\
&=& \frac{m_X^2}{2m_e T + m_X^2} \varepsilon^{V(A)}_{e\ell^\prime}~,
\label{eq::coup_tilde_nls1b_VA}
\end{eqnarray}
where $\varepsilon^{V(A)}_{e\ell^\prime}$ can be defined as
\begin{eqnarray}
\varepsilon^V_{e\ell^\prime}&=&\frac{\varepsilon^L_{e\ell^\prime}
+ \varepsilon^R_{e\ell^\prime}}{2}
= \frac{V_{e\ell^\prime}}{2\sqrt2 G_{F}m_X^2} \nonumber \\
\varepsilon^A_{e\ell^\prime}&=&\frac{\varepsilon^L_{e\ell^\prime}
- \varepsilon^R_{e\ell^\prime}}{2}
=\frac{A_{e\ell^\prime}}{2\sqrt2 G_{F}m_X^2}~.
\label{eq::coup_nls1b_VA}
\end{eqnarray}

\begin{table*} [hbt]
\caption{
Constraints at 90\% C.L. on the couplings for the FV NLS1B with $m_X^2\ll2m_eT$ and
$m_X^2\gg2m_eT$ for the TEXONO and LSND data sets
obtained from a one-parameter-at-a-time analysis.}
\label{tab::nls1b_fv}
\begin{ruledtabular}
\begin{tabular}{cccc}
\multicolumn{2}{c}{($m_X^2\ll2m_eT$)} & \multicolumn{2}{c}{($m_X^2\gg2m_eT$)} \\ \ \\
TEXONO PC-Ge $(\times 10^{-6})$ & LSND $(\times 10^{-6})$ &
TEXONO CsI(Tl) $(\times 10^{-6})$ & LSND $(\times 10^{-6})$ \\ \hline \\

${|L_{e\mu(\tau)}|}^{1/2} < 0.94$ & ${|L_{e\mu(\tau)}|}^{1/2} < 19.15$ &
${|L_{e\mu(\tau)}|}^{1/2}/m_X < 5.16$ & ${|L_{e\mu(\tau)}|}^{1/2}/m_X < 3.63$ \\ \ \\

${|R_{e\mu(\tau)}|}^{1/2} < 0.94$ & ${|R_{e\mu(\tau)}|}^{1/2} < 30.75$ &
${|R_{e\mu(\tau)}|}^{1/2}/m_X < 2.57$ & ${|R_{e\mu(\tau)}|}^{1/2}/m_X < 6.48$ \\ \ \\

${|V_{e\mu(\tau)}|}^{1/2} < 0.79$ & ${|V_{e\mu(\tau)}|}^{1/2} < 18.58$ &
${|V_{e\mu(\tau)}|}^{1/2}/m_X < 2.59$ & ${|V_{e\mu(\tau)}|}^{1/2}/m_X < 3.56$ \\ \ \\

${|A_{e\mu(\tau)}|}^{1/2} < 0.79$ & ${|A_{e\mu(\tau)}|}^{1/2} < 18.50$ &
${|A_{e\mu(\tau)}|}^{1/2}/m_X < 2.48$ & ${|A_{e\mu(\tau)}|}^{1/2}/m_X < 3.55$ \\ \ \\

\end{tabular}
\end{ruledtabular}
\end{table*}

The differential cross sections as a function of the
recoil energy $T$ with typical reactor $\nuebar$ spectra
for the NLS1B at some specific coupling and mass parameters
using the Ge detector as a target for both the FC and FV cases are
displayed in Fig.~\ref{fig::diff_nls1b} for illustration,
where the SM contribution is superimposed. As can be seen in this figure,
the cross section shows different behavior with respect to the recoil
energy $T$, which provides more advantages in the measurements of
the couplings at low energy for the low mass values of $m_X$.
Because of the $1/T$ dependency in the cross section, working at
low-energy threshold provides better sensitivity in the
coupling for small mass values of $m_X$. The $2m_eT$ term in
the denominator can be safely neglected for high values of $m_X$.
Therefore, the CsI(Tl) detector data are expected to provide more
stringent limits since the cross section is measured at a good
sensitivity in the 3-8 MeV range.

\begin{figure*}[]
  \begin{center}
  {\bf \hspace{1.cm}(a)} {\bf \hspace{8.5cm}(b)}\\
    \includegraphics[width=8.5cm]{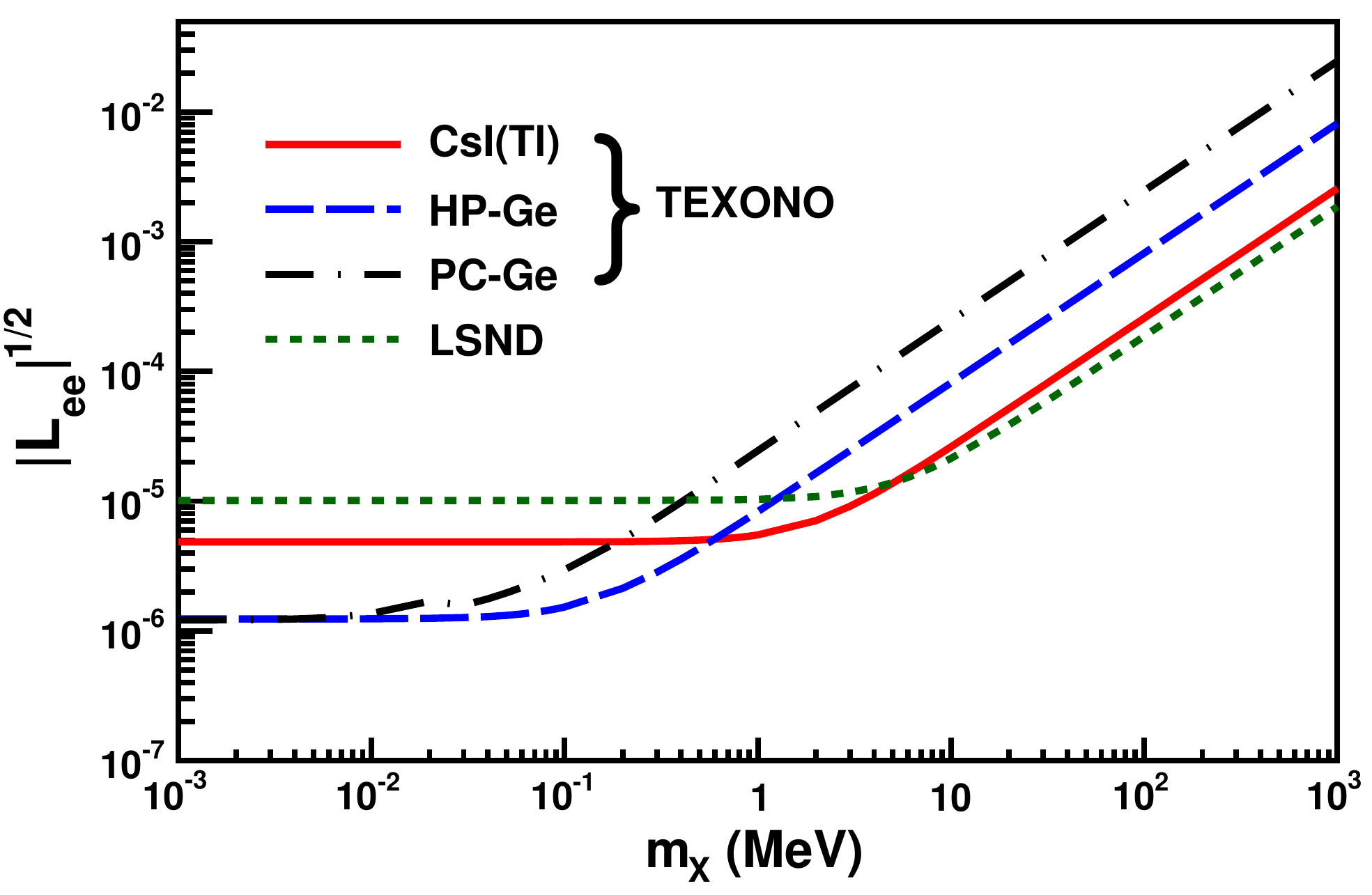} \qquad
    \includegraphics[width=8.5cm]{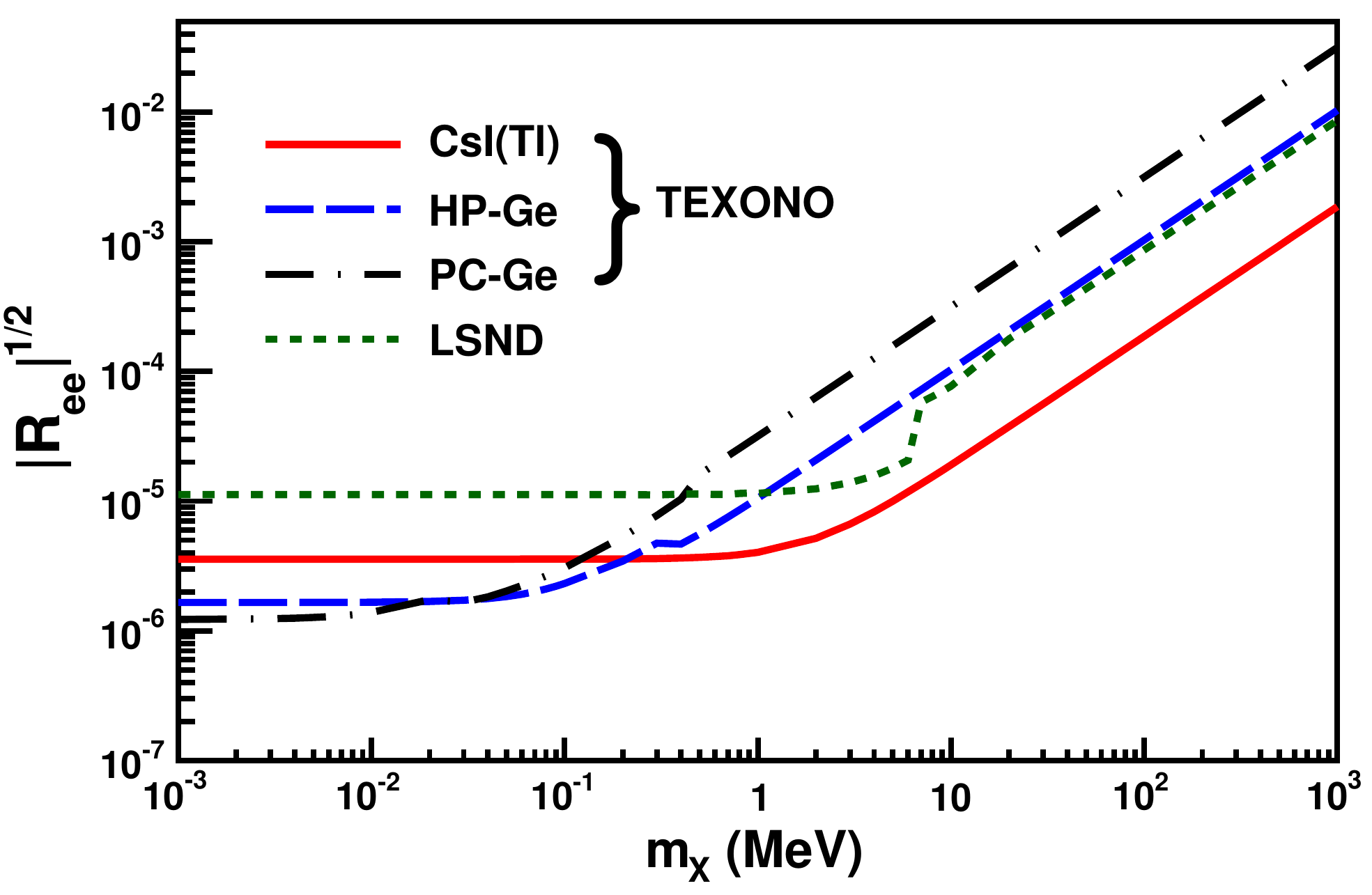}\\
  {\bf \hspace{1.cm}(c)} {\bf \hspace{8.5cm}(d)}\\
    \includegraphics[width=8.5cm]{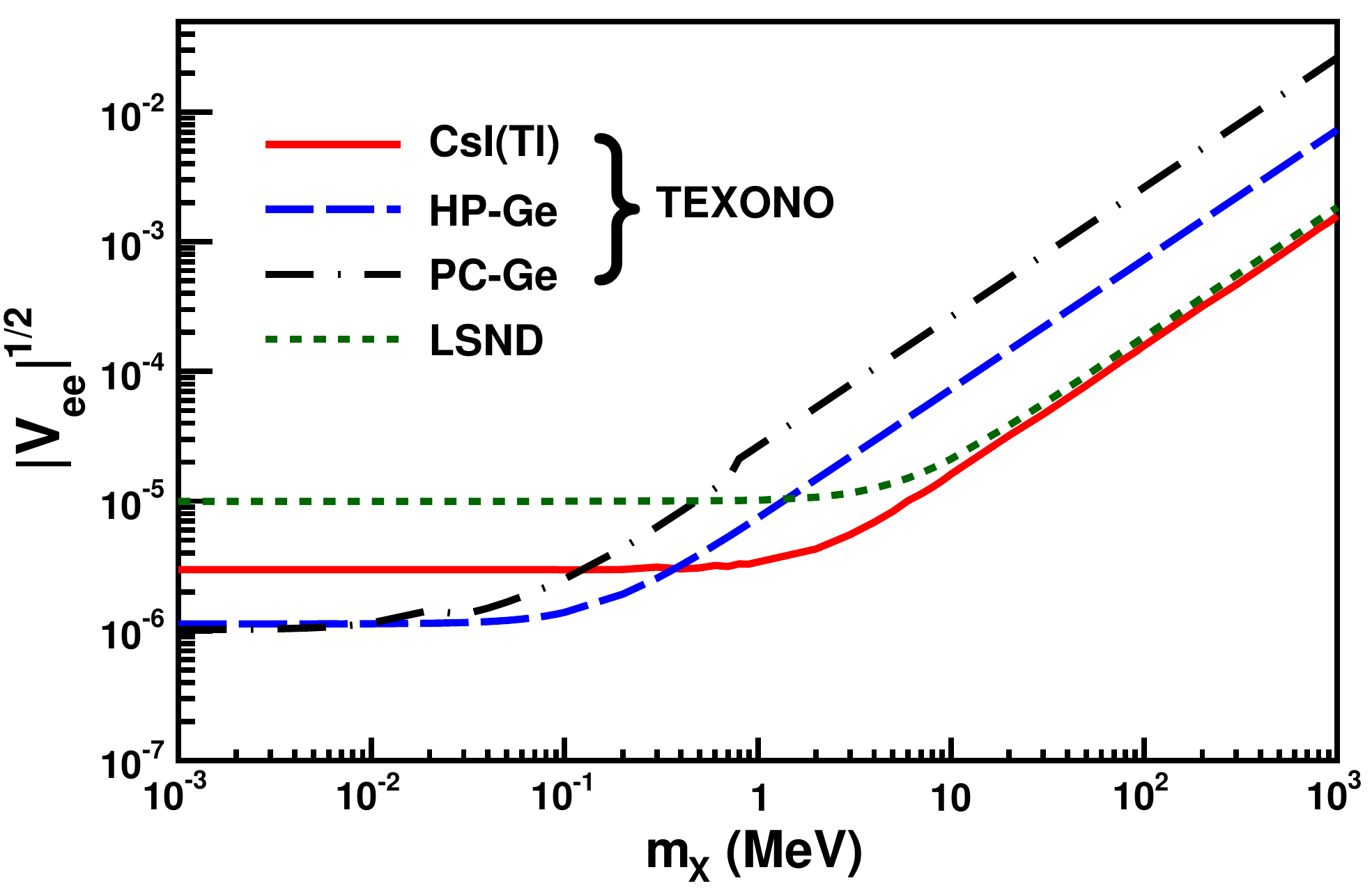} \qquad
    \includegraphics[width=8.5cm]{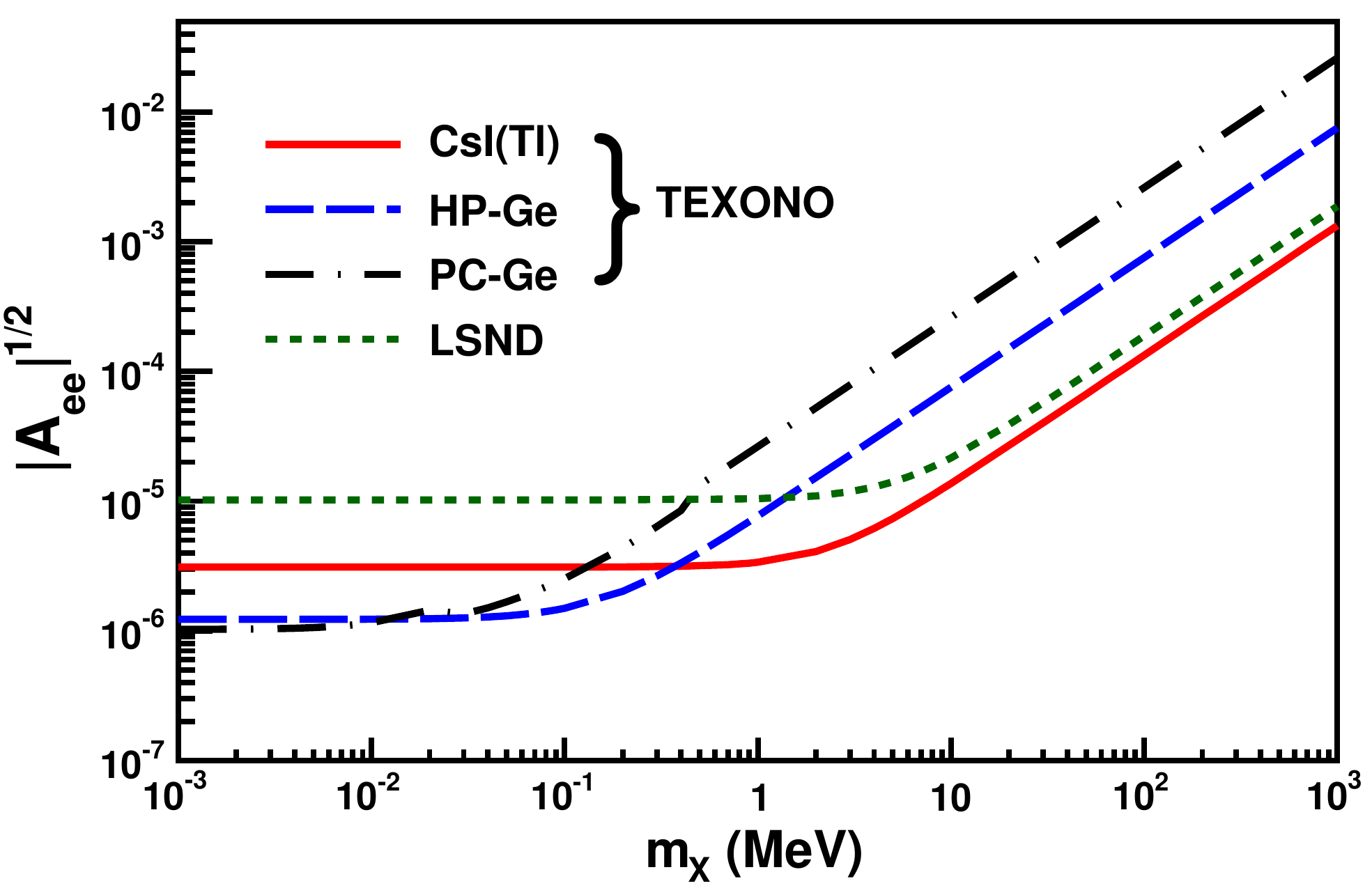}\\
  {\bf \hspace{1.cm}(e)} {\bf \hspace{8.5cm}(f)}\\
    \includegraphics[width=8.5cm]{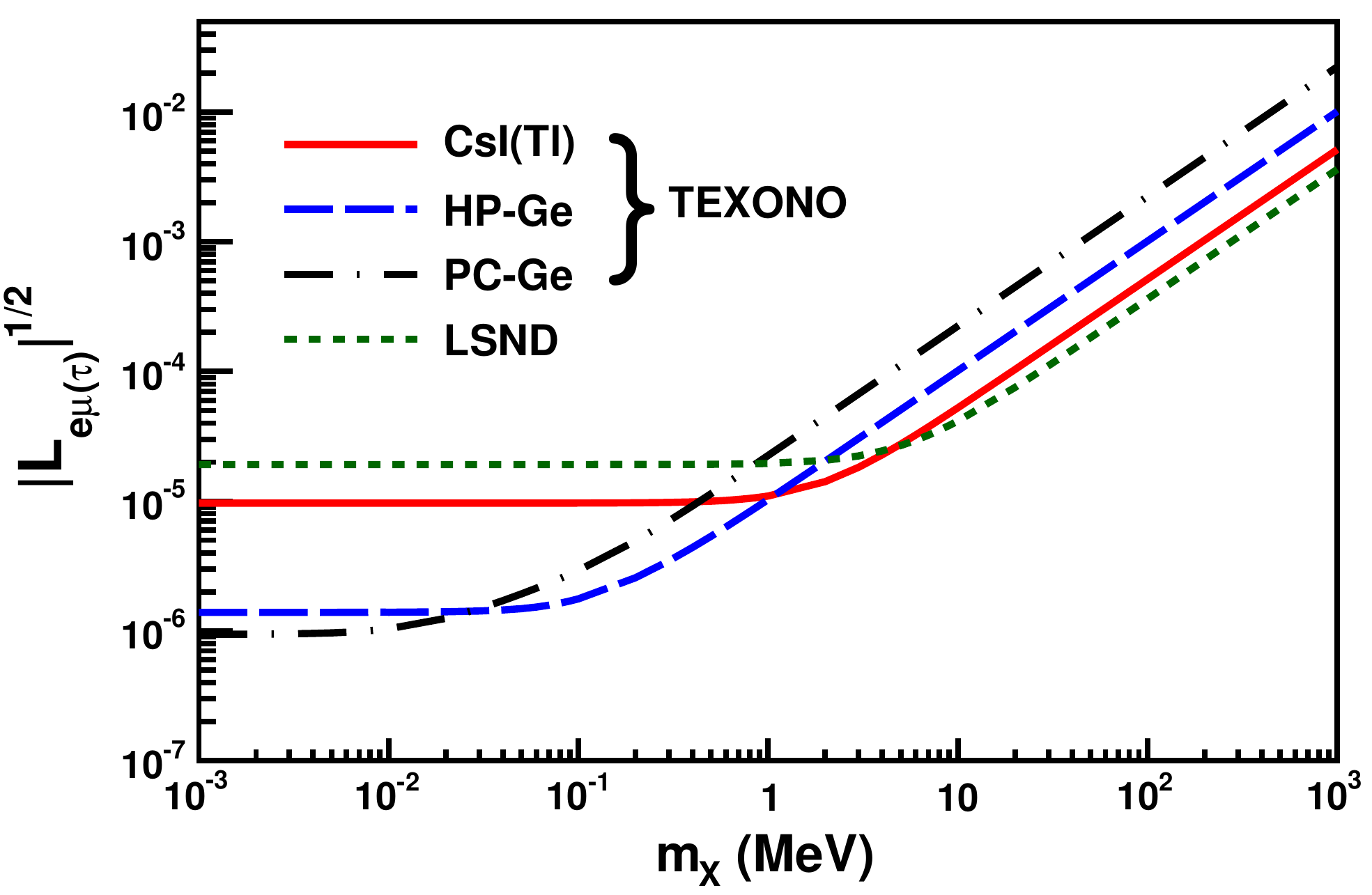} \qquad
    \includegraphics[width=8.5cm]{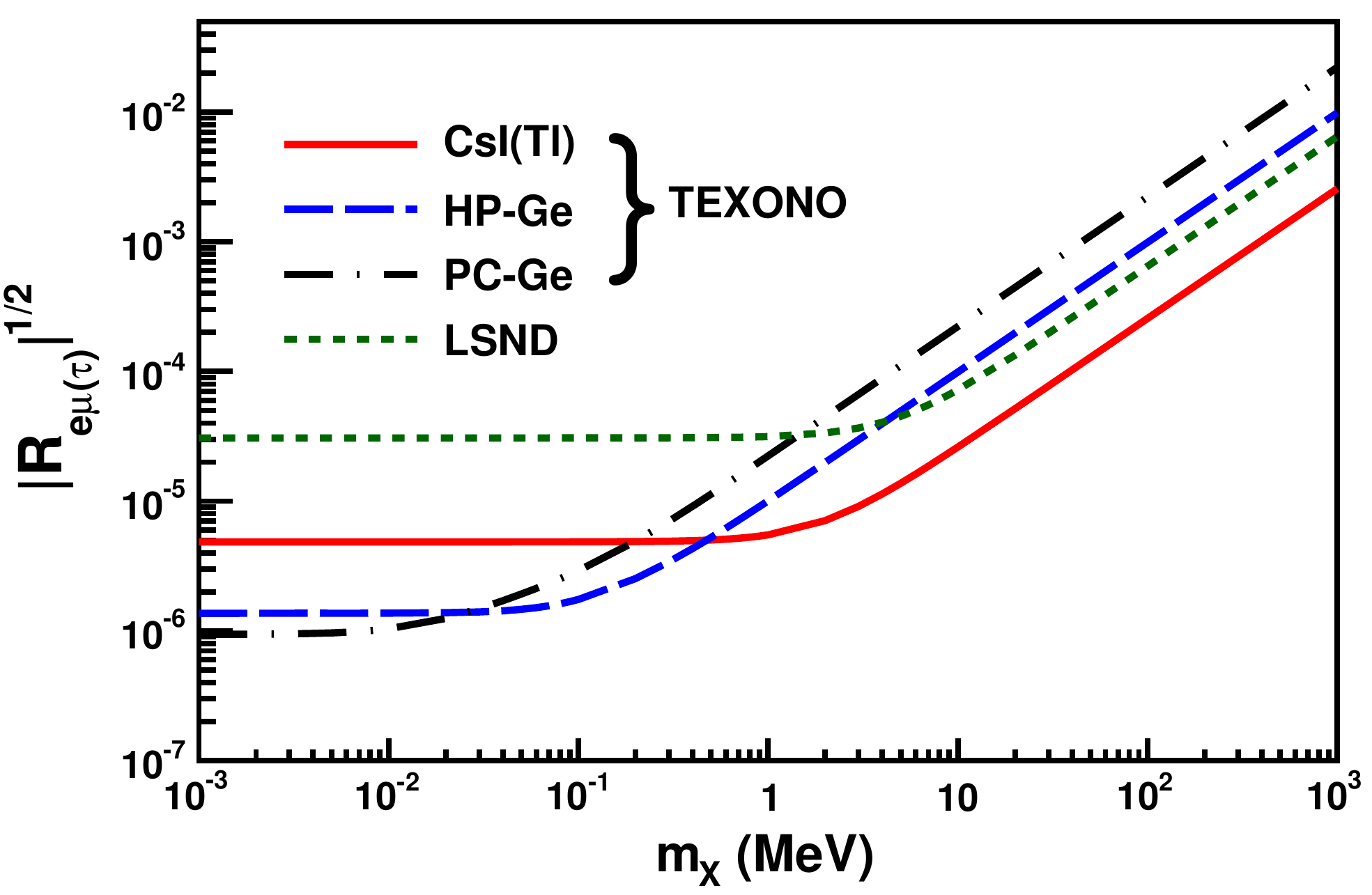}\\
  {\bf \hspace{1.cm}(g)} {\bf \hspace{8.5cm}(h)}\\
    \includegraphics[width=8.5cm]{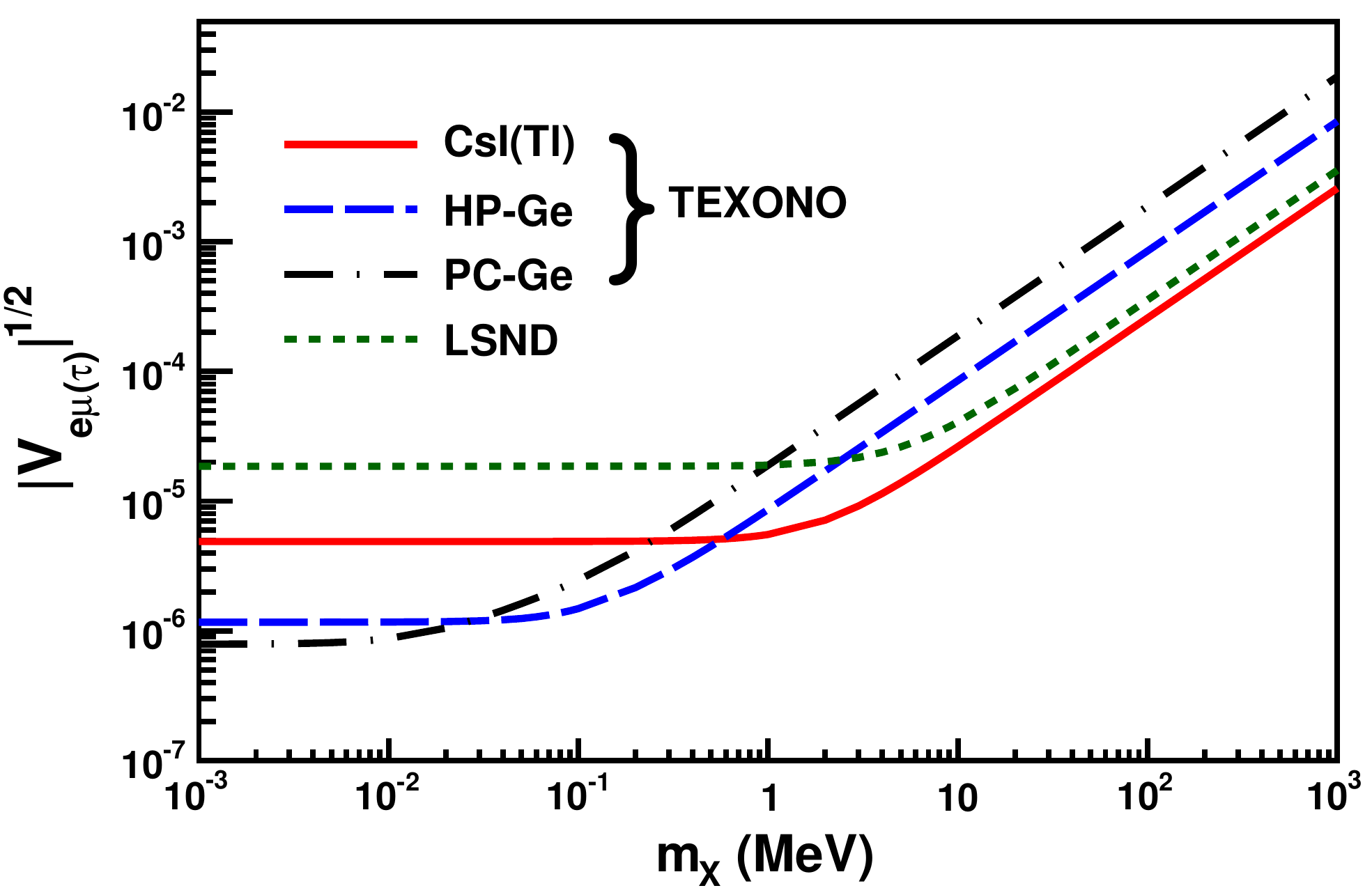} \qquad
    \includegraphics[width=8.5cm]{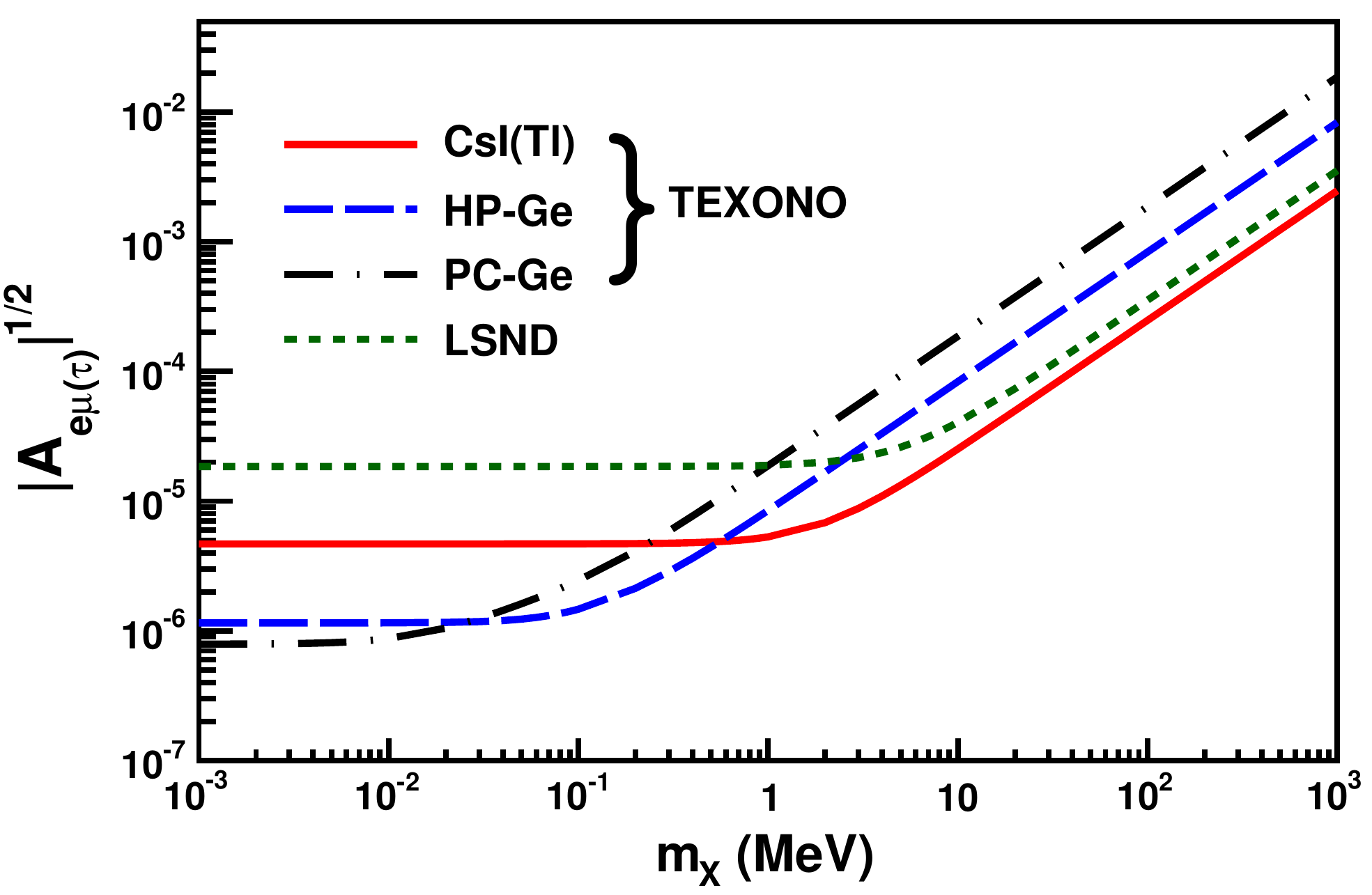}\\
    \caption{The 90\% C.L. upper limits for
    the couplings of (a) $|L_{ee}|^{1/2}$, (b) $|R_{ee}|^{1/2}$,
    (c) $|V_{ee}|^{1/2}$, (d) $|A_{ee}|^{1/2}$,
    (e) $|L_{e\mu(\tau)}|^{1/2}$, (f) $ |R_{e\mu(\tau)}|^{1/2}$,
    (g) $|V_{e\mu(\tau)}|^{1/2}$, and (h) $|A_{e\mu(\tau)}|^{1/2}$
    for TEXONO and LSND with various $m_{X}$ values
    by adopting a one-parameter-at-a-time analysis.}
    \label{fig::s1_res_lr_va}
  \end{center}
\end{figure*}

\begin{figure*}[]
  \begin{center}
  {\bf \hspace{1.cm}(a)} {\bf \hspace{7.5cm}(b)}\\
    \includegraphics[width=7cm]{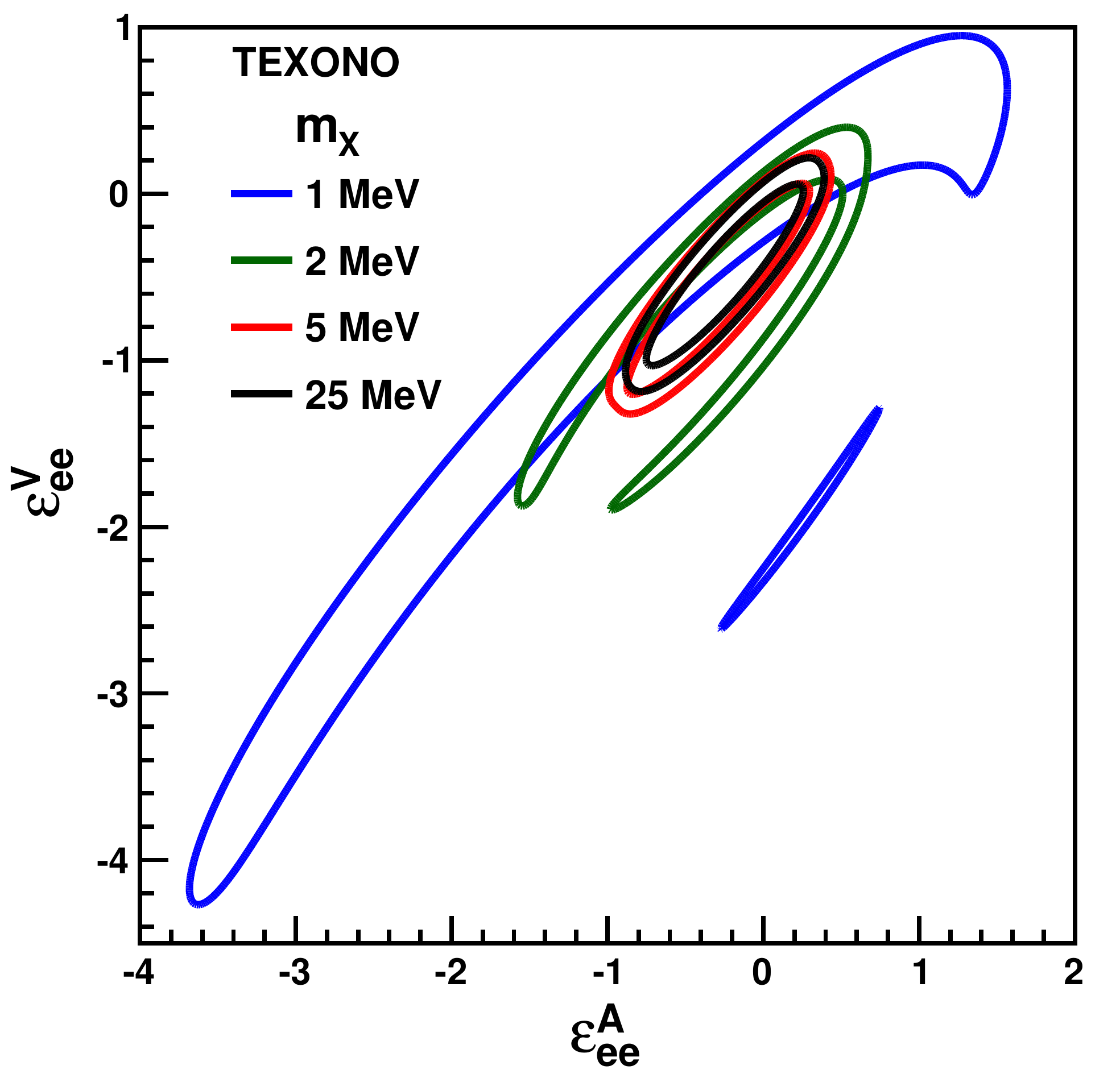} \qquad
    \includegraphics[width=7cm]{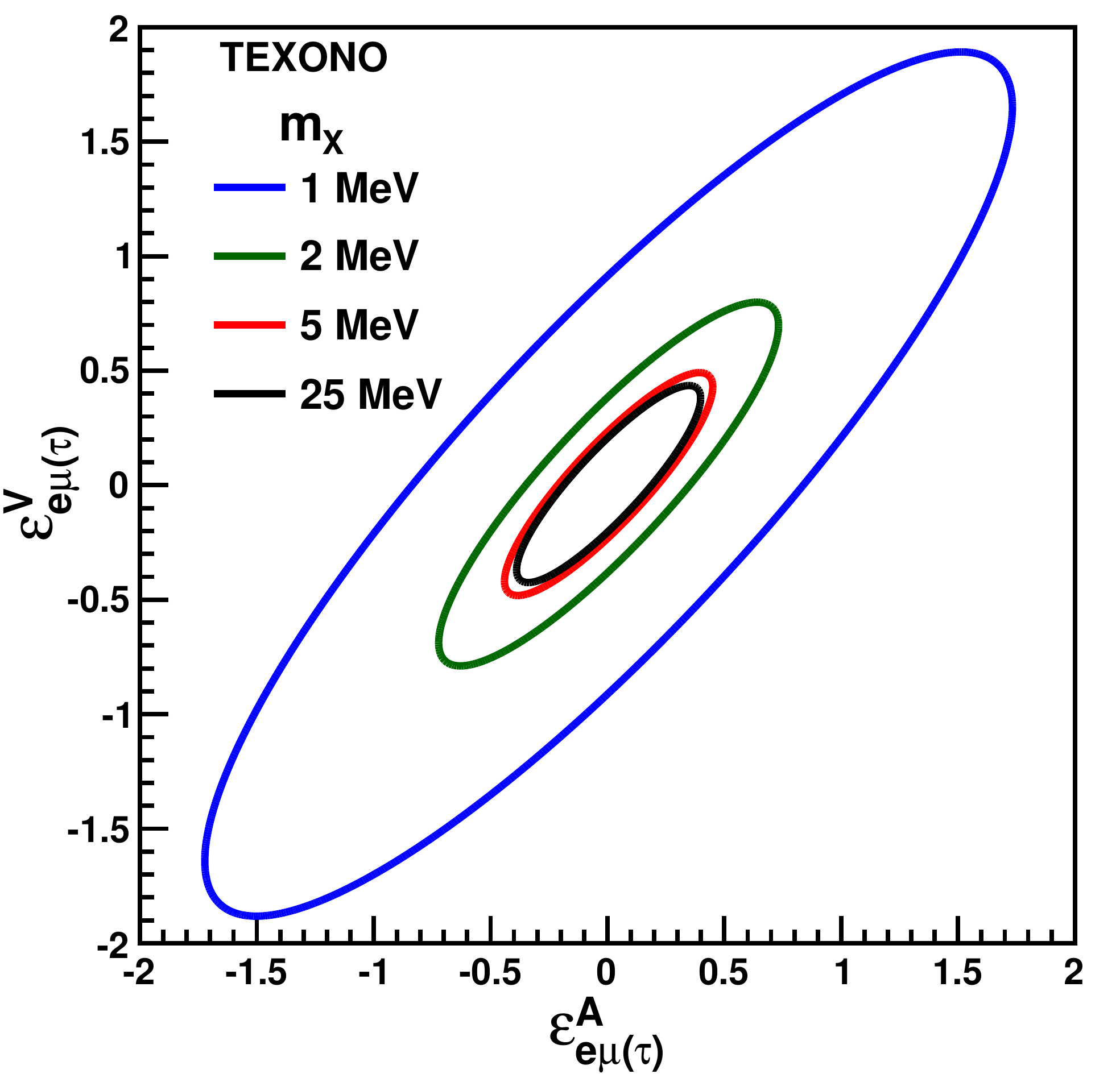} \\
  {\bf \hspace{1.cm}(c)} {\bf \hspace{7.5cm}(d)}\\
    \includegraphics[width=7cm]{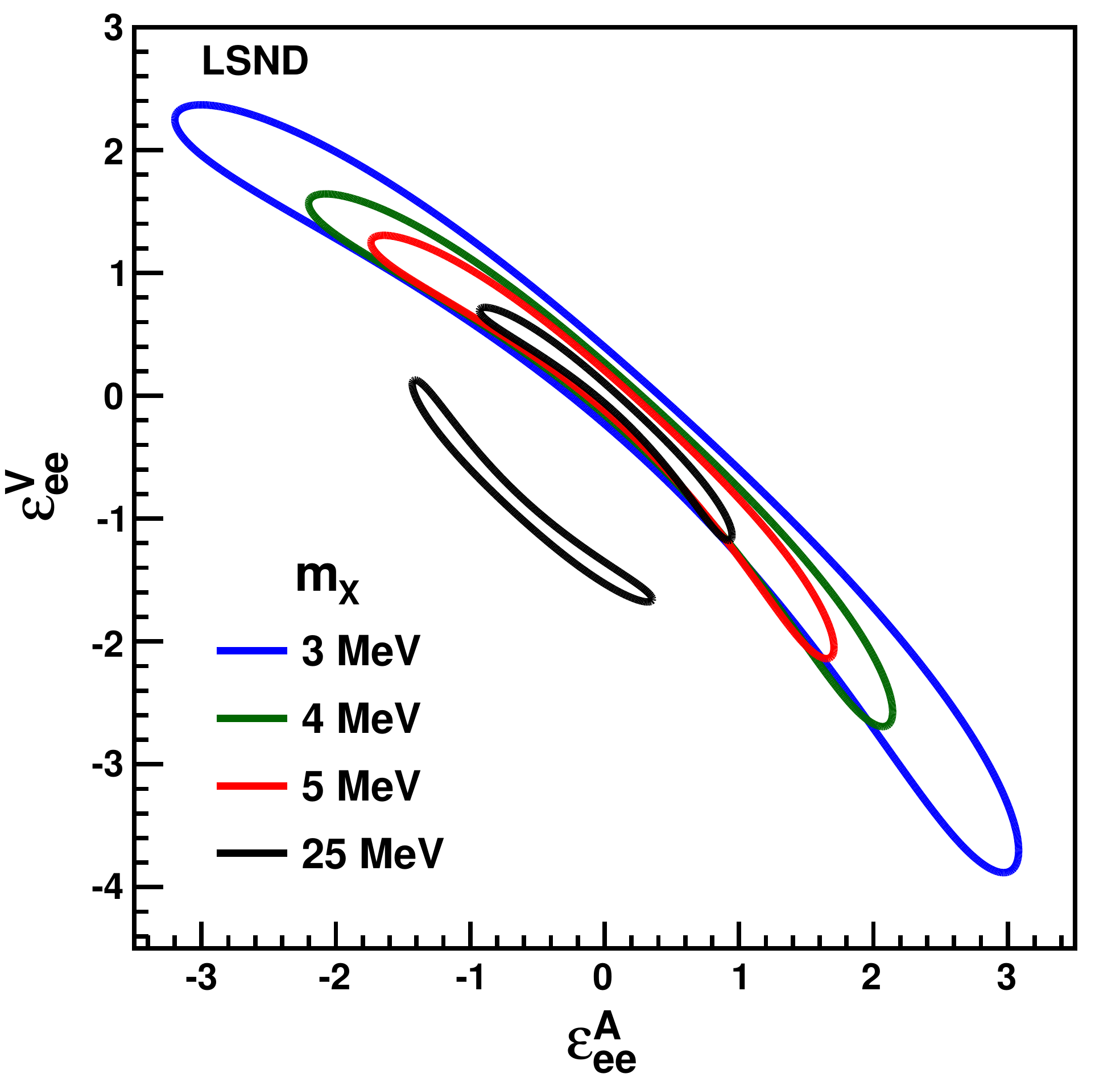} \qquad
    \includegraphics[width=7cm]{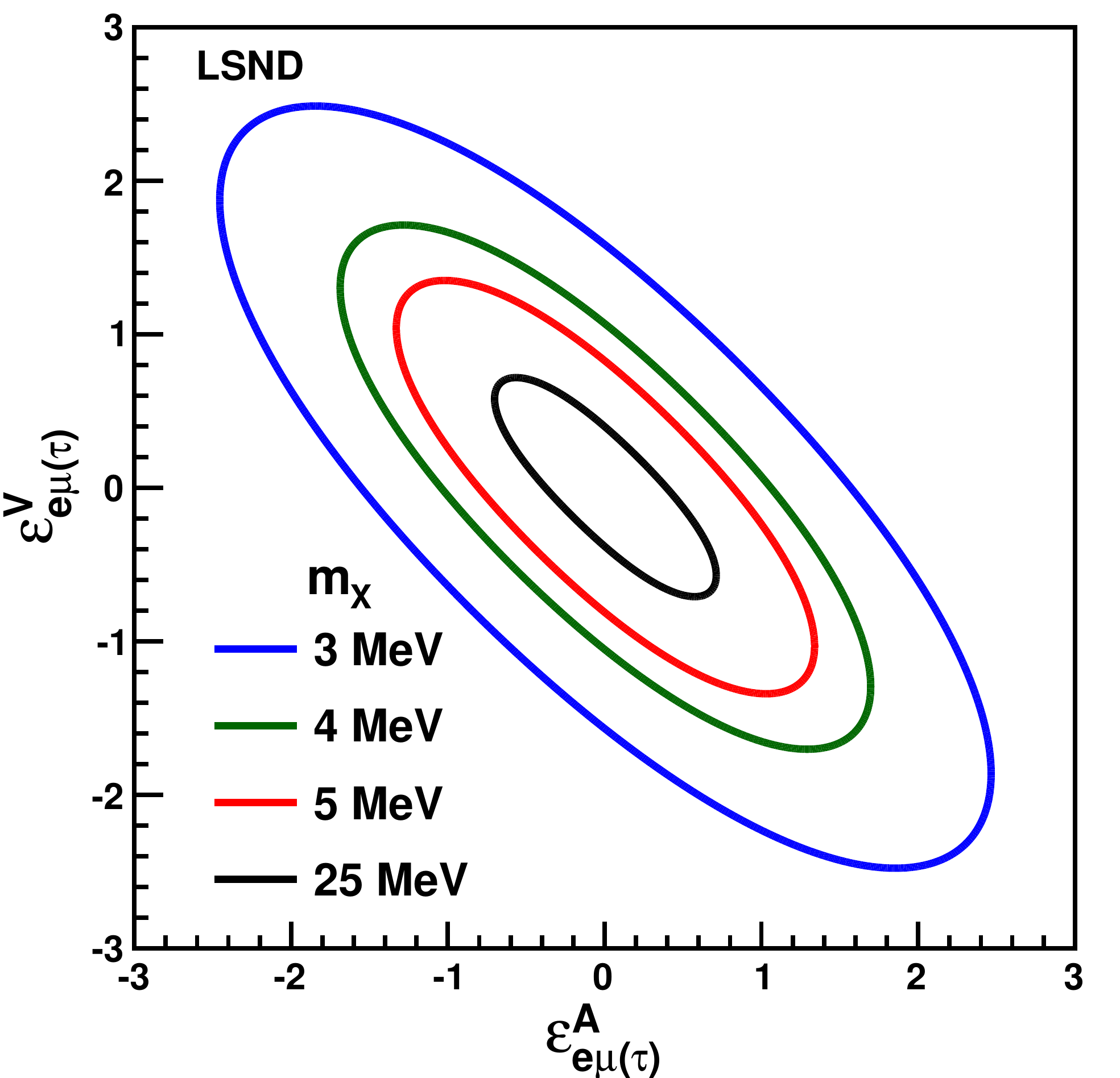}\\
  {\bf \hspace{1.cm}(e)} {\bf \hspace{7.5cm}(f)}\\
    \includegraphics[width=7cm]{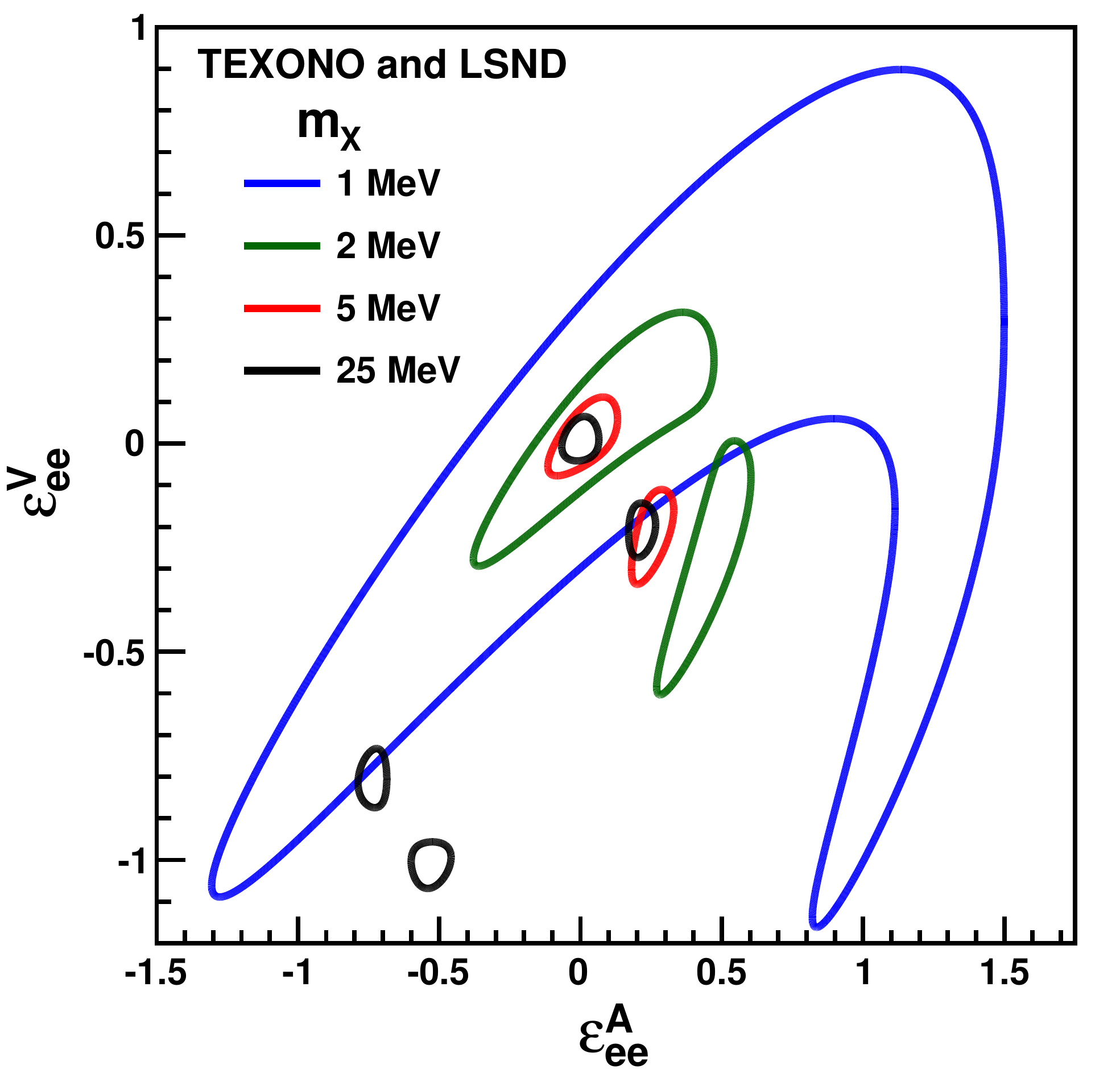} \qquad
    \includegraphics[width=7cm]{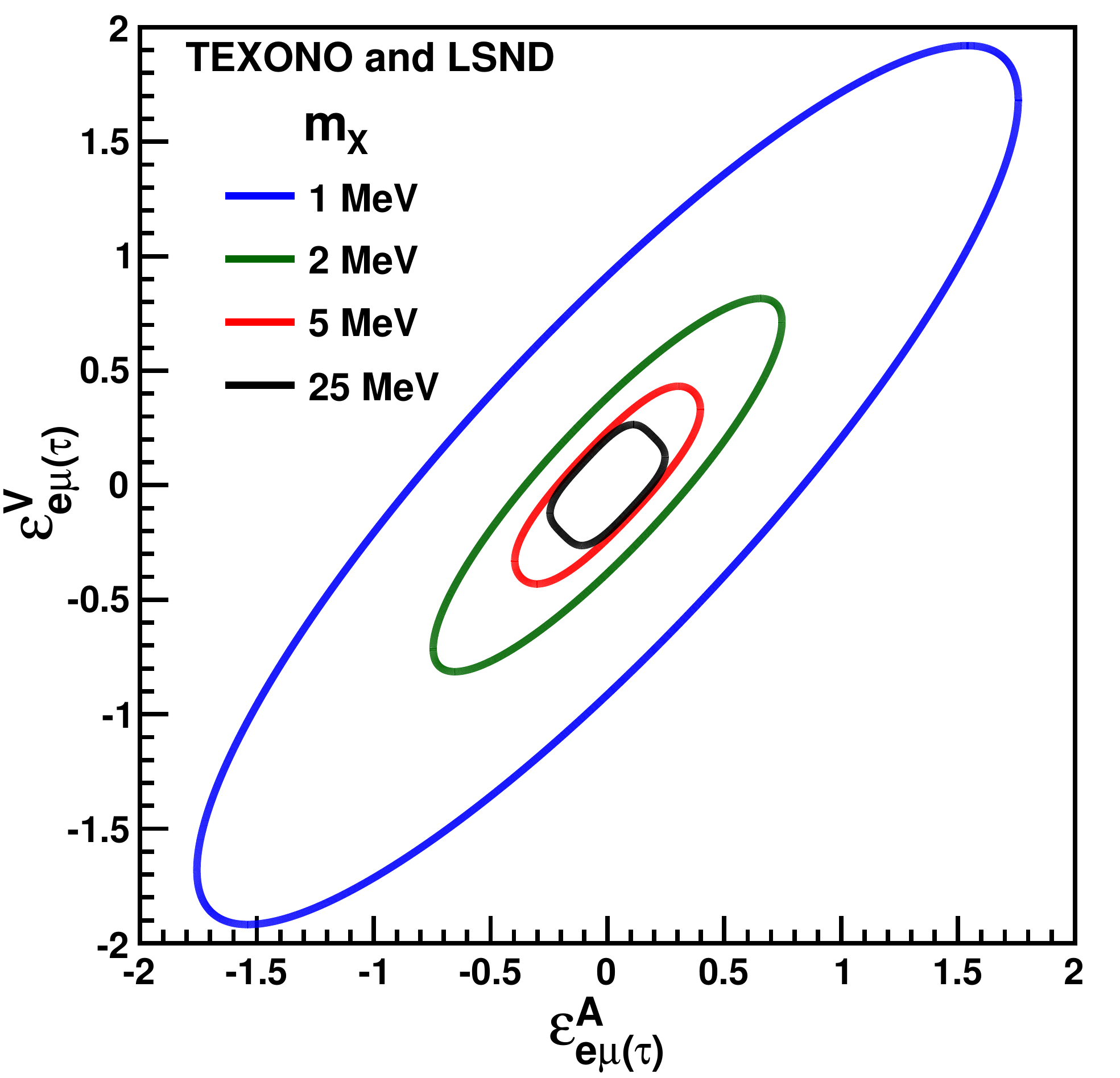} \\
    \caption{The allowed regions at 90\% C.L. for
    (a) the FC NLS1B in the parameter space of $\varepsilon^V_{ee}$
    and $\varepsilon^A_{ee}$; (b) FV NLS1B in the parameter space of $\varepsilon^V_{e\mu(\tau)}$
    and $\varepsilon^A_{e\mu(\tau)}$ for TEXONO CsI(Tl) with various $m_{X}=$ 1, 2, 5, 25 MeV, from
    outer to inner, respectively; (c) FC NLS1B in the parameter space of $\varepsilon^V_{ee}$
    and $\varepsilon^A_{ee}$; (d) FV NLS1B in the parameter space of $\varepsilon^V_{e\mu(\tau)}$
    and $\varepsilon^A_{e\mu(\tau)}$ for LSND with various $m_{X}=$ 3, 4, 5, 25 MeV, from
    outer to inner, respectively, with the global fitting for allowed regions of TEXONO CsI(Tl)
    and LSND at 90\% C.L. for (e) FC NLS1B couplings of $\varepsilon^V_{ee}$ vs $\varepsilon^A_{ee}$,
    and (f) FV NLS1B couplings of $\varepsilon^V_{e\mu(\tau)}$
    vs $\varepsilon^A_{e\mu(\tau)}$ with various $m_{X}=$ 1, 2, 5, 25 MeV from
    outer to inner, respectively.}
    \label{fig::s1_va}
  \end{center}
\end{figure*}

In particular, the NLS1B cross section has
$1/(2m_eT+m_X^2)$ dependency, which is directly proportional to the
sensitivity of the couplings of $L(R)_{e\ell^\prime}$.
When $T$ and $m_X$ become comparable, the $2m_eT$ term in the denominator
cannot be neglected anymore. This term, however, causes us to lose the sensitivity
in $\varepsilon^{L(R)}$ for low mass values of $m_X$.
When $m_X$ gets bigger, i.e., $m_X \gtrsim 25$ MeV,
$2m_eT$ can be neglected and the sensitivity stays fixed
at the minimum value as shown in Fig.~\ref{fig::s1_res}.
In this case, the $\tilde{\varepsilon}^{L(R)}_{e\ell^\prime} =
\varepsilon^{L(R)}_{e\ell^\prime}$ and
$\tilde{\varepsilon}^{V(A)}_{e\ell^\prime}=
\varepsilon^{V(A)}_{e\ell^\prime}$ conditions
are satisfied.

For the FC NLS1B interaction, the allowed regions at 90\% C.L.
in the parameter space of $\varepsilon^L_{ee}$ and $\varepsilon^R_{ee}$
with various $m_{X}=$ 1, 2, 5, 25 MeV for TEXONO and $m_{X}=$ 3, 4, 5, 25 MeV for
LSND are illustrated in Figs.~\ref{fig::s1_res}(a) and
\ref{fig::s1_res}(c), respectively.

Similarly, for the FV NLS1B interaction, the allowed regions at 90\% C.L. for
the couplings of $\varepsilon^L_{e\mu(\tau)}$ and $\varepsilon^R_{e\mu(\tau)}$
with various $m_{X}=$ 1, 2, 5, 25 MeV for TEXONO and $m_{X}=$ 3, 4, 5, 25 MeV for
LSND are illustrated in Figs.~\ref{fig::s1_res}(b) and
\ref{fig::s1_res}(d), respectively.

The global fitting for allowed regions
of TEXONO and LSND for the couplings of $\varepsilon^L_{e\ell\prime}$ and
$\varepsilon^R_{e\ell\prime}$ at 90\% C.L. with various $m_{X}=$ 1, 2, 5, 25 MeV
are illustrated in Figs.~\ref{fig::s1_res}(e) and \ref{fig::s1_res}(f)
for FC and FV NSI, respectively.

By adopting a one-parameter-at-a-time analysis in the minimum $\chi^2$ analysis,
the bounds at 90\% C.L. on the FC and FV NLS1B
couplings for low and high mass values are given in
Tables~\ref{tab::nls1b_fc} and \ref{tab::nls1b_fv}, and the upper limits
at 90 \% C.L. are illustrated in Fig.~\ref{fig::s1_res_lr_va} with respect
to mass parameter $m_X$. As shown in Table~\ref{tab::nls1b_fc},
Table~\ref{tab::nls1b_fv}, and Fig.~\ref{fig::s1_res_lr_va}, the TEXONO
PC-Ge and HP-Ge data provide better constraints in ${L(R)}_{e\ell^\prime}$
and ${V(A)}_{e\ell^\prime}$ parameter spaces compared to LSND for both FC
and FV NLS1B in the case of $m_X \ll 2m_eT$. On the other hand, TEXONO CsI(Tl)
gives better constraints in the $L(R)_{e\ell^\prime}$ and ${V(A)}_{e\ell^\prime}$
parameter spaces compared to LSND for both FC and FV NLS1B
in the case of $m_X \gg 2m_eT$.

The 90\% C.L. upper limits for the couplings of $|L_{ee}|^{1/2}$, $|R_{ee}|^{1/2}$
and $|L_{e\mu(\tau)}|^{1/2}$, $|R_{e\mu(\tau)}|^{1/2}$ vs mass parameter of
$m_{X}$ for TEXONO and LSND are illustrated in
Figs.~\ref{fig::s1_res_lr_va}(a) and \ref{fig::s1_res_lr_va}(b) and
Figs.~\ref{fig::s1_res_lr_va}(e) and \ref{fig::s1_res_lr_va}(f), respectively.
The 90\% C.L. upper limits for the couplings of $|V_{ee}|^{1/2}, |A_{ee}|^{1/2}$
and $|V_{e\mu(\tau)}|^{1/2}$, $|A_{e\mu(\tau)}|^{1/2}$
versus mass parameter of $m_{X}$ for TEXONO CsI(Tl) and LSND are
illustrated in Figs.~\ref{fig::s1_res_lr_va}(c) and ~\ref{fig::s1_res_lr_va}(d) and
Figs.~\ref{fig::s1_res_lr_va}(g) and ~\ref{fig::s1_res_lr_va}(h), respectively.

Similarly, the allowed regions at 90\% C.L. in the parameter space
of $\varepsilon^{V}_{ee}$ and $\varepsilon^{A}_{ee}$ with various $m_X$
for TEXONO CsI(Tl) and LSND are illustrated in Figs.~\ref{fig::s1_va}(a) and
\ref{fig::s1_va}(c), respectively, for the FC NLS1B.
In the case of the FV NLS1B, the allowed regions at 90\% C.L. in the parameter space of
$\varepsilon^{V}_{e\mu(\tau)}$ and $\varepsilon^{A}_{e\mu(\tau)}$
with various $m_X$ for TEXONO CsI(Tl) and LSND are illustrated in
Figs.~\ref{fig::s1_va}(b) and \ref{fig::s1_va}(d), respectively.

The global fitting for allowed regions
of TEXONO and LSND for the couplings of $\varepsilon^{V}_{e\ell\prime}$ and
$\varepsilon^{A}_{e\ell\prime}$ at 90\% C.L. with various $m_{X}=$ 1, 2, 5, 25 MeV
are illustrated in Figs.~\ref{fig::s1_va}(e) and \ref{fig::s1_va}(f)
for FC and FV NSI, respectively.

\begin{figure}[!ht]
  \begin{center}
  {\bf (a)}\\
  \includegraphics[height=5.cm]{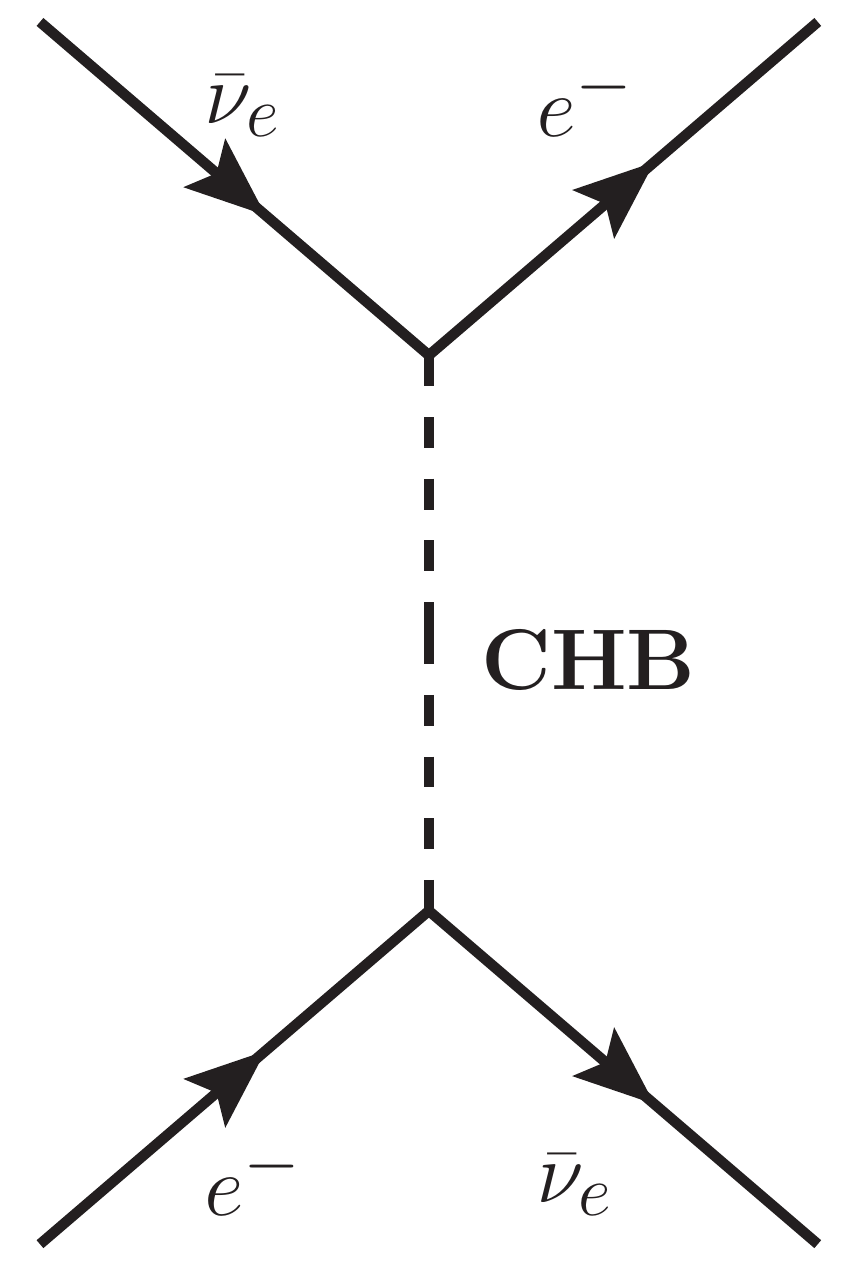}\\
  {\bf (b)}\\
  \includegraphics[width=6.cm]{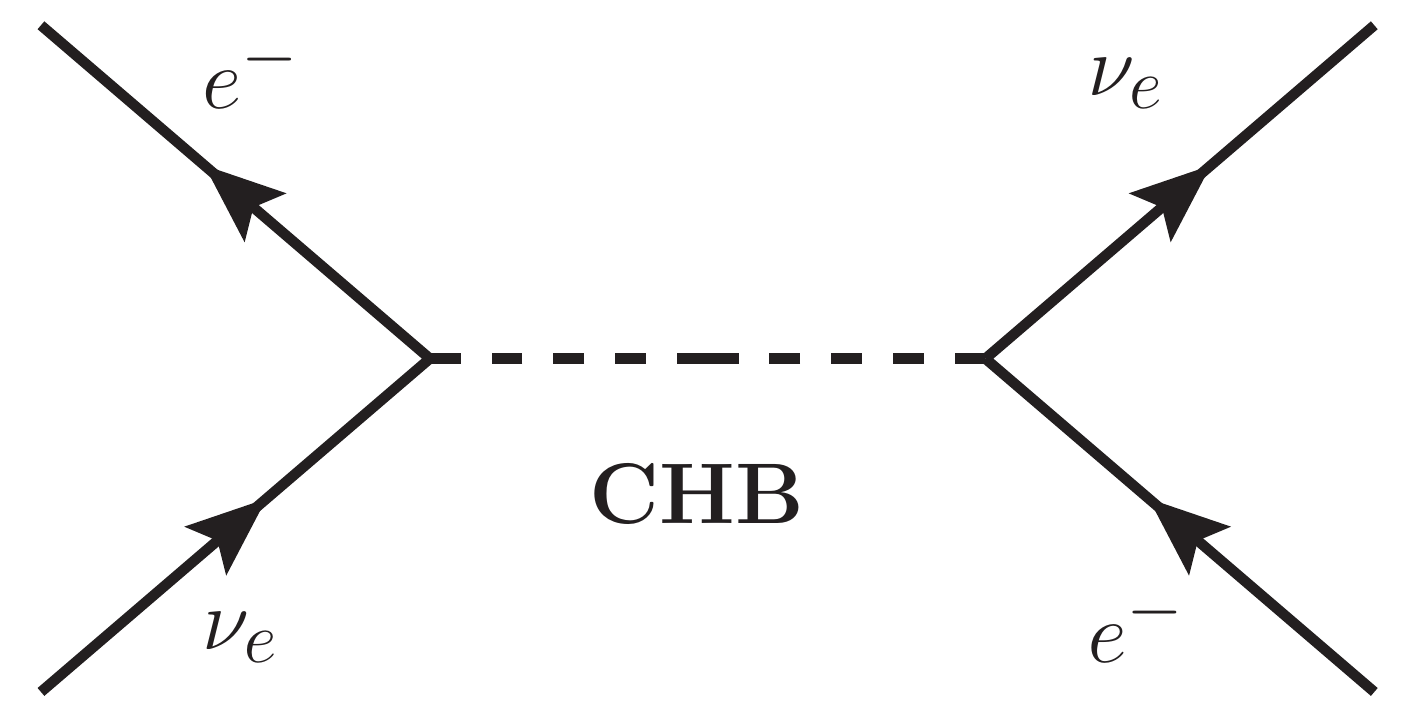}
  \caption{\label{fig::feyn_chb} Feynman diagrams of
  (a) $\nuebare$  scattering and
  (b) $\nuee$ scattering for the exchange of charged Higgs boson.}
  \end{center}
\end{figure}

\begin{table} [!hbt]
\caption{Upper bounds at 90\% C.L. on the coupling of $h_{ee}$ in CHB interaction
for $\nuebare$ and $\nuee$ scattering.}
\label{tab::chb}
\begin{ruledtabular}
\begin{tabular}{ccc}

\multirow{ 2}{*}{$M_H$} & \multicolumn{2}{c}{$h_{ee}$ $(\times10^{-6})$} \\

& {TEXONO}  & {LSND} \\ \hline \\

1 MeV & $<$ 4.99 & $<$ 8.03 \\ \ \\

2 MeV & $<$ 0.09 & $<$ 2.91 \\ \ \\

2.2 MeV & $<$ 0.02 & $<$ 1.22 \\ \ \\

2.5 MeV & $<$ 0.06 & $<$ 0.32 \\ \ \\

2.9 MeV & $<$ 0.02 & $<$ 2.88 \\ \ \\

3 MeV & $<$ 4.58 & $<$ 1.47 \\ \ \\

4 MeV & $<$ 8.45 & $<$ 0.46 \\ \ \\

5 MeV & $<$ 11.46 & $<$ 2.26 \\ \ \\

6 MeV &  $<$ 14.27 & $<$ 17.67 \\ \ \\

10 MeV & $<$ 25.02 & $<$ 57.57 \\

\end{tabular}
\end{ruledtabular}
\end{table}

\begin{figure}[!ht]
  \begin{center}
  \includegraphics[width=8.5cm]{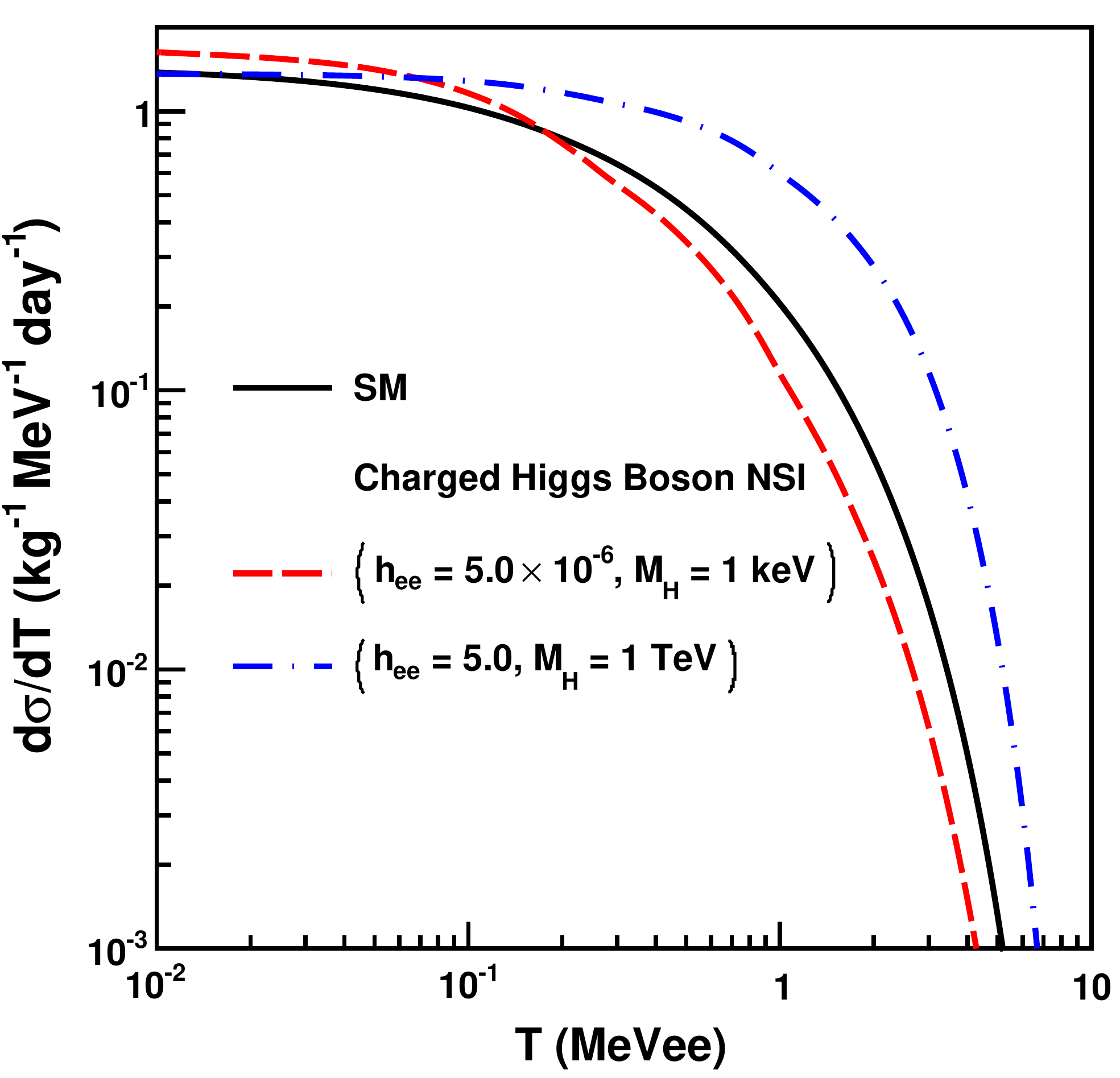}\\ \ \\
  \caption{\label{fig::diff_chb} Differential cross section
  as a function of recoil energy $T$ with a typical reactor $\nuebar$ spectrum
  for the exchange of CHB. The SM differential cross section
  is superimposed for comparison.}
  \end{center}
\end{figure}

\subsection{Charged Higgs boson}

Leptons, quarks and gauge bosons acquire their mass through the Higgs mechanism
~\cite{Englert64}, while neutrinos still remain massless in the SM.
In order to introduce and explain the smallness of neutrino masses without
requiring an extra right-handed neutrino, one of the simplest models among other
mechanisms is the Higgs triplet model (HTM), through which neutrinos gain their
mass~\cite{Cheng80,Ong13}. In HTM, apart from the neutral scalar Higgs boson ($h^0$),
there also appear singly charged ($H^{+}$) and doubly charged ($H^{++}$) ones,
since Higgs triplets under the standard $SU(2)_L$ gauge group have two units
of weak hypercharge.

There are many phenomenological studies at high-energy accelerator experiments
such as LHC and Tevatron in the literature ~\cite{Fileviez08,Akeroyd11}.
However, in this study we also consider the low-energy frontier with
$\nuebare$ and $\nuee$ elastic scattering, which are pure leptonic processes
providing an elegant test to the SM of electroweak theory.
The Feynman diagrams of $\nuebare$ and $\nuee$ scattering via the exchange of CHB
are displayed in Fig.~\ref{fig::feyn_chb}.

In the HTM, the electroweak $\rho$ parameter is predicted at the tree level as
$\rho \simeq 1-2 v^2_{\triangle}/v^2_{\Phi}$, where $v_{\Phi}$ and $v_{\triangle}$
are the vacuum expectation values of the doublet Higgs
field and triplet Higgs field, respectively.  However, the experimental value
of this parameter $\rho_{exp}=1.0004^{+0.0003}_{-0.0004}$~\cite{shinya12}
requires that $v_{\triangle}$ be smaller than a few GeV,
i.e., $v_{\triangle}<$ 3.5~GeV at \%95 C.L., and
hence $v_{\triangle}/v_{\Phi} \lesssim 0.02 $.
Taking these into account,  the interaction Lagrangian
for the coupling of the CHB to leptons can be written as
\begin{equation}\label{eq::lag_chb}
 {\cal L} = -h_{\ell\ell'}\sqrt{2}(\ell^T C P_L \nu_{\ell'}+
 \nu^T_\ell C P_L \ell')H^+ + H.c.,
\end{equation}
where $h_{\ell\ell'}$ is the coupling constant;
$\ell(\ell') = $ e, $\mu$, or $\tau$; $C$ is the charge conjugation;
and $P_L$ is the chiral projector.

\begin{figure*}[!ht]
  \begin{center}
  {\bf \hspace{1.cm}(a)} {\bf \hspace{8.cm}(b)}\\
    \includegraphics[width=7.75cm]{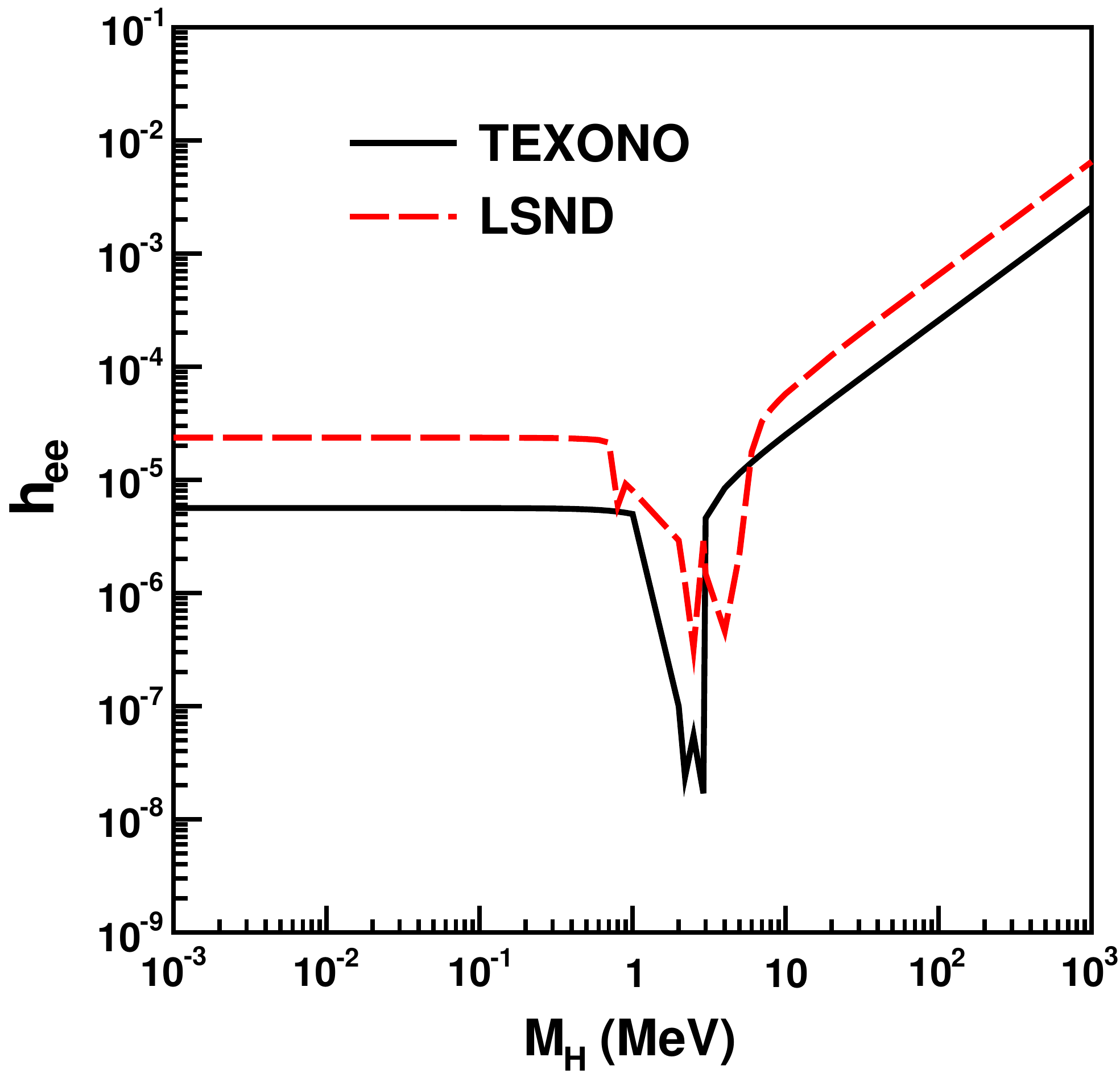}\qquad
    \includegraphics[width=7.75cm]{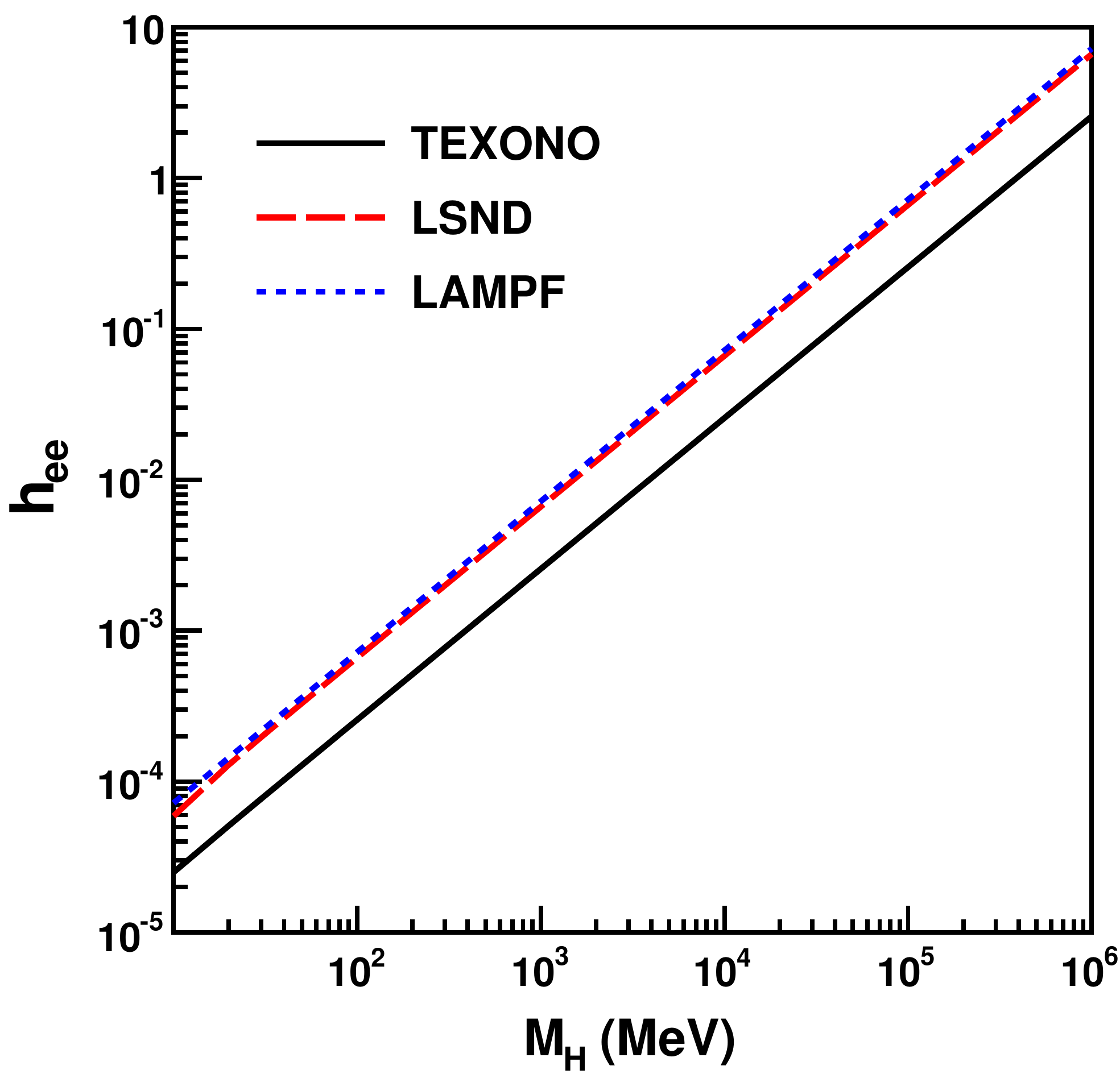}
    \caption{The upper limit of coupling $h_{ee}$
    with respect to  the mass of CHB $M_H$ at 90\% C.L. for
    (a) low and (b) high mass values.}
    \label{fig::chb_MH_hee}
  \end{center}
\end{figure*}

The $\nuebare$ and $\nuee$ scattering differential cross sections for
CHB exchange contributions are found, respectively, to be
\begin{equation}\label{eq::dif_chb}
\left[\frac{d\sigma_{\nuebar e}}{dT}\right]_{\text{CHB}} =
\frac{m_e}{4\pi} \frac{[h_{ee}^2]^2}{[m_e(m_e+2E_\nu)-M_H^2]^2}
\end{equation}
and
\begin{equation}
\left[\frac{d\sigma_{\nue e}}{dT}\right]_{\text{CHB}} =
\frac{m_e}{4\pi} \frac{[h_{ee}^2]^2(1-T/E_\nu)^2}
{[m_e^2+2m_e(E_\nu-T)-M_H^2]^2}.
\label{eq::dif_neu}
\end{equation}

The differential cross section for CHB
with relevant parameters for TEXONO CsI(Tl)
is displayed in Fig.~\ref{fig::diff_chb} with different
mass parameters, where SM contribution
is superimposed for comparison.

For the high mass value of CHB, the terms of $m_e(m_e+2E_\nu)$
or $ m_e^2+2m_e(E_\nu-T)$ in the denominator can be neglected.
Therefore, $(h_{ee}/M_H)^4$ becomes a fitting parameter.
From the best fit,
\begin{equation}
(h_{ee}/M_H)^4 = [8.32 \pm 16.74 \pm 13.39] \times 10^{-12} ~\text{GeV}^{-4}
\label{eq::chb_fit_texo1}
\end{equation}
is obtained at $\chi^2/dof = 8.8/9 $ for TEXONO CsI(Tl) data. Similarly,
\begin{equation}
(h_{ee}/M_H)^4 = [5.21 \pm 6.10 \pm 4.51] \times 10^{-10} ~\text{GeV}^{-4}
\label{eq::chb_fit_lsnd1}
\end{equation}
is obtained at $\chi^2/dof = 9.7/13 $ for LSND. They can be converted to their
corresponding upper limit at 90\% C.L. of
\begin{equation}
h_{ee}/M_H < 2.57 \times 10^{-3} ~\text{GeV}^{-1}
\label{eq::chb_limit_texo1}
\end{equation}
for TEXONO CsI(Tl) and
\begin{equation}
h_{ee}/M_H < 6.48 \times 10^{-3} ~\text{GeV}^{-1}
\label{eq::chb_limit_lsnd1}
\end{equation}
for LSND. TEXONO provides more stringent limits than
those from LSND and a previous study given in Ref.~\cite{coarasa96}
as $h_{ee}/M_H < 7.2 \times 10^{-3} ~ GeV^{-1}$
for the LAMPF $\nuee$ experiment~\cite{allen93} at 90\% C.L.,
which was derived based on the measurement value
of $sin^2\theta_W$.

On the other hand, for the low mass value of CHB,
the $M_H$ term in the denominator can be neglected. Therefore,
only $(h_{ee})^4$ becomes a fitting parameter. From the best fit,
\begin{equation}
(h_{ee})^4 = [1.10 \pm 4.07 \pm 3.65] \times 10^{-22}
\label{eq::chb_fit_texo2}
\end{equation}
is obtained at $\chi^2/dof = 8.8/9 $ with its corresponding
upper limit at 90\% C.L. of
\begin{equation}
h_{ee} < 5.63 \times 10^{-6}
\label{eq::chb_limit_texo2}
\end{equation}
for TEXONO.

Similarly, for LSND, from the best fit,
\begin{equation}
(h_{ee})^4 = [6.59 \pm 11.51 \pm 9.35] \times 10^{-20}
\label{eq::chb_fit_lsnd2}
\end{equation}
is obtained at $\chi^2/dof = 10.3/13 $ with its corresponding
upper limit at 90\% C.L. of
\begin{equation}
h_{ee} < 23.6 \times 10^{-6}
\label{eq::chb_limit_lsnd2}
\end{equation}
for the low mass value of CHB.

The upper limit of coupling $h_{ee}$ with respect to the CHB
mass values of $M_H $ for TEXONO CsI(Tl) and LSND at 90\% C.L.
for low and high mass values are shown in Figs.~\ref{fig::chb_MH_hee}(a)
and \ref{fig::chb_MH_hee}(b), respectively. Upper bounds at 90\% C.L. on the
coupling of $h_{ee}$ for $ \bar{\nu}_{e}-e $ and $ \nu_{e}-e $ scattering
for various mass values of $M_H$ are listed in Table~\ref{tab::chb}.
In this study, the parameter space is extended and consequently a new window
is opened for the low mass CHB.

\section{SUMMARY and PROSPECTS}

In summary, in this article, some of the BSM new physics
scenarios including massive intermediate particles such as the
NLS1B, $Z'$, and CHB have been discussed and their potential to explain
some of the anomalous effects that cannot be explained by SM has been addressed.

The experimental results of upper bounds for NSI using data from
the analysis of the $\nuebare$ and $\nuee$ elastic scattering interaction
cross section measurements were placed in the framework of these
BSM scenarios. The existing experimental sensitivities were improved,
and the parameter space was extended by including the low-energy regime.

Particularly, in the NLS1B study, a new research window has been opened
for a low mass NLS1B in the low-energy regime due to $1/T$ dependency in the
cross section. For a low mass NLS1B, the coupling becomes directly
proportional to $1/T$; therefore, working at low energy and low threshold
becomes substantially  important to see the effect of BSM.
In this study we found that TEXONO gives better constraints in
$\varepsilon^{L(R)}_{e\ell^\prime}$ parameter space compared to
the neutrino-electron channel, i.e., the LSND and LAMPF  $\nuee$ experiments,
for both the FC and FV NLS1B cases.

In the literature, many studies on high energies have targeted
high mass values of CHB. However, in this study we also considered the
low-energy frontier with $\nuebare$ and $\nuee$ elastic scattering, which
are pure leptonic processes providing an elegant test of the
electroweak theory of SM. We have found new limits on the CHB couplings
with respect to mass covering the low mass CHB region.

In our study of $Z'$, the current limits were not improved since
the experimental uncertainties are big compared to
the heavy expectation value of the $Z'$ mass. However, it is still interesting
enough to look for the mass limits of $Z'$ at the low-energy, low-momentum regime.
This study showed that if the experimental uncertainties were improved by 1\%,
the current existing limits could be reached via the neutrino-electron scattering channel.
By the help of the projection, it is possible to investigate the relationship between
the $Z'$ mass bounds and experimental accuracies that may provide intuitive scaling for
future neutrino experiments.

\section{ACKNOWLEDGMENTS}

This work is supported by Contract No. 114F374 under the Scientific and Technological
Research Council of Turkey (T\"{U}B\.ITAK); Contract No.
104-2112-M-001-038-MY3 from the Ministry of Science and Technology, Taiwan;
and Contract No. 2017-ECP2 from the National Center of Theoretical Sciences, Taiwan.

\end{document}